\documentclass{article}
\usepackage{arxiv}

\usepackage[utf8]{inputenc} % allow utf-8 input
\usepackage[T1]{fontenc}    % use 8-bit T1 fonts
\usepackage{hyperref}       % hyperlinks
\usepackage{url}            % simple URL typesetting
\usepackage{booktabs}       % professional-quality tables
\usepackage{amsfonts}       % blackboard math symbols
\usepackage{nicefrac}       % compact symbols for 1/2, etc.
\usepackage{microtype}      % microtypography
\usepackage{lipsum}		% Can be removed after putting your text content
\usepackage{graphicx}
\usepackage{doi}

\usepackage{subfigure}
\usepackage{booktabs}
\usepackage{stackengine}
\usepackage{mathtools}
\usepackage{tikz}
\usepackage{array}
\title{Snapshot Ptychography on Array Cameras}

\author{ 
    \href{https://orcid.org/0000-0002-6910-3978}{\includegraphics[scale=0.06]{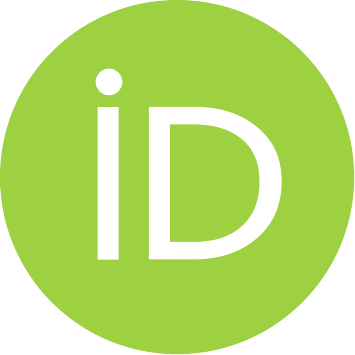}\hspace{1mm}Chengyu Wang}\thanks{Chengyu Wang and Minghao Hu are students at Duke University. This work was finished when they were doing internship at the University of Arizona.}\\
	Wyant College of Optical Sciences\\
	University of Arizona\\
	Tucson, AZ 85721 \\
	\texttt{chengyuwang@arizona.edu} \\
	\And
	{Minghao Hu} \\
	Wyant College of Optical Sciences\\
	University of Arizona\\
	Tucson, AZ 85721 \\
	\texttt{mh432@arizona.edu} 
	\And
	\href{https://orcid.org/0000-0003-3011-9320}{\includegraphics[scale=0.06]{orcid.pdf}\hspace{1mm}Yuzuru Takashima} \\
	Wyant College of Optical Sciences\\
	University of Arizona\\
	Tucson, AZ 85721 \\
	\texttt{ytakashima@optics.arizona.edu} 
	\And
	{Timothy J. Schulz} \\
	Department of Electrical and Computer Engineering\\
	Michigan Technological University\\
	Houghton, MI 49931 \\
	\texttt{schulz@mtu.edu} 
	\And
	\href{https://orcid.org/0000-0001-5655-2478}{\includegraphics[scale=0.06]{orcid.pdf}\hspace{1mm}David J. Brady} \\
	Wyant College of Optical Sciences\\
	University of Arizona\\
	Tucson, AZ 85721 \\
	\texttt{djbrady@arizona.edu} 
}

\begin{document}
\maketitle

\begin{abstract}
We use convolutional neural networks to recover images optically down-sampled by $6.7\times$ using coherent aperture synthesis over a 16 camera array. Where conventional ptychography relies on scanning and oversampling, here we apply decompressive neural estimation to recover full resolution image from a single snapshot, although as shown in simulation multiple snapshots can be used to improve SNR. In place training on experimental measurements eliminates the need to directly calibrate the measurement system. We also present simulations of diverse array camera sampling strategies to explore how snapshot compressive systems might be optimized. 
\end{abstract}

%%%%%%%%%%%%%%%%%%%%%%%%%%  body  %%%%%%%%%%%%%%%%%%%%%%%%%%
\section{Introduction}

The cross range resolution of diffractive imaging systems is aperture limited. Radar imaging has long used aperture synthesis from moving or distributed receivers to increase aperture~\cite{ryle1960synthesis}. Numerous studies of synthetic aperture ladar have attempted to extend this advantage to optical frequencies~\cite{beck2005synthetic}. While such systems have been demonstrated with significant range and resolution~\cite{krause2011synthetic, wang2018inverse}, the challenges of holographic stability and referencing have limited their applicability. Recently, reference-free aperture synthesis using ptychography has been increasingly popular, beginning with the seminal demonstration by Zheng {\it et al.} of a gigapixel-scale microscope~\cite{zheng2013wide}. While gigapixel-scale aperture synthesis has also been demonstrated by holographic methods~\cite{brady2011gigapixel, fienup2011gigapixel}, Fourier ptychography (FP) requires no reference signal and was implemented by Zheng {\it et al.} with a simple LED illumination array. 

Diverse approaches have subsequently been proposed to improve the resolution~\cite{ou2015high}, portability~\cite{dong2014fpscope}, or acquisition speed~\cite{bian2014content} of this setup. Single-shot FP has been demonstrated with a diffractive grating~\cite{he2018single}, a lens array~\cite{lee2018single}, or color multiplexing~\cite{sun2018single}. Multi-camera systems capture band-limited images in parallel using multiple cameras to increase the imaging throughput~\cite{chan2019parallel,kim2016incubator, konda2021multi}. Aperture-scanning FP translates the aperture with a mechanical stage~\cite{dong2014aperture} or performs digital scanning with a spatial light modulator (SLM)~\cite{ou2016aperture}. The development of aperture-scanning FP further permits macroscopic super-resolution imaging where far-field propagation is equivalent to the Fourier transform of the target field~\cite{dong2014aperture,holloway2016toward,holloway2017savi}. Beyond the increased resolution and space-bandwidth product, the advantages of FP also include phase imaging~\cite{ou2013quantitative}, digital refocusing~\cite{dong2014aperture}, 3D imaging~\cite{horstmeyer2016diffraction} and aberration correction~\cite{chung2016wide}. Recent comprehensive reviews of FP are presented in Konda {\em et al.}~\cite{konda2020fourier} and Zheng {\em et al.}~\cite{zheng2021concept}.

Here we consider the extension of array camera aperture synthesis to macroscopic imaging systems. Such an extension was previously suggested by Holloway {\em et al.}~\cite{holloway2016toward}, but to our knowledge has yet to be demonstrated. Calibration of the forward mapping from object space to the sensor array is the primary challenge to such a demonstration. We use data measured directly on the array to train a convolutional neural network to directly invert multiaperture data, avoiding both the calibration problem and reliance on phase retrieval algorithms. In so doing, we also demonstrate single frame synthetic aperture imaging, which ultimately may enable video-rate multiaperture coherent imaging. Similar methods have previously been demonstrated in diverse applications of snapshot compressive imaging~\cite{yuan2021snapshot}.

Conventional Fourier ptychography uses phase retrieval algorithms, as reviewed for example in~\cite{yeh2015experimental} to combine coherent image data across multiple frames. Phase retrieval algorithms depend on data redundancy; typical systems require at least 60\% overlap in the Fourier space between adjacent images~\cite{holloway2016toward}. The scanning associated redundant sampling and the iterative nature of the reconstruction algorithms lead to long acquisition and processing times. To address the processing aspect of this challenge, deep-learning (DL)-based algorithms have been proposed~\cite{jiang2018solving,schulz2021photon}. Kappeler {\it et al.}~\cite{kappeler2017ptychnet} proposed a 3-layer CNN and demonstrated reconstruction performance better than alternating projection methods when there was no overlap. Nguyen {\it et al.}~\cite{nguyen2018deep} proposed a conditional generative adversarial network (cGAN) and reported 40 times faster reconstruction. Boominathan {\it et al.}~\cite{boominathan2018phase} proposed a U-Net with different training strategies for high overlap and low overlap cases, and showed improved reconstruction in all cases. Shamshad {\it et al.}~\cite{shamshad2019deep} utilized generative models with subsampling operator which required less observed data and was more robust to noise. To improved the network generalization, Zhang {\it et al.}~\cite{zhang2019fourier} proposed to synthesize a complex field from the measurements as the input to the network. Xue {\it et al.}~\cite{xue2019reliable} proposed to reconstruct the phase and assess the estimated phase at the same time using a Bayesian convolutional neural network.

While simulations and experiments have demonstrated that DL methods improve the imaging efficiency in FP, less attention has been paid to the their advantages in hardware design. In fact, because DL reconstruction need not rely on the analytic forward model, accurate system calibration is no longer needed. Here we show that a fixed array camera can be used with end-to-end neural training to recover images upscaled up by 6.7x in resolution relative to the single aperture limit. Our contributions include demonstration of an "in place" training strategy and testing strategies to confirm synthetic aperture performance. In particular, we use data selection to show that observed resolution enhancements are intrinsically tied to array size. In addition to our experimental demonstration, we present simulations of diverse array sampling strategies, including multiframe strategies based on subaperture array translation. 

While innovations in calibration strategy and processing beyond the scope of this study will be needed to field coherent multiaperture cameras, the results presented here confirm the ability of such systems to greatly exceed the single aperture diffraction limit and the utility of neural processing in image formation from such systems. 
 
Section~\ref{sec:proto} describes the experimental system we built to demonstrate the proposed approach. Section~\ref{sec:inverse} details our neural training and estimation strategy and experimental results. Section~\ref{sec:simulations} presents design simulations to help understand the impact of subaperture size and distribution and compression ratio on system performance. Finally section~\ref{sec:conclusion} discussed the significance of results presented here and potential next steps in the development of coherent array imaging.

\section{System Design}
\label{sec:proto}
\begin{figure}[htbp]
\subfigure[Array camera snapshot FP\label{fig:scheme_our}]{\includegraphics[width=0.5\textwidth]{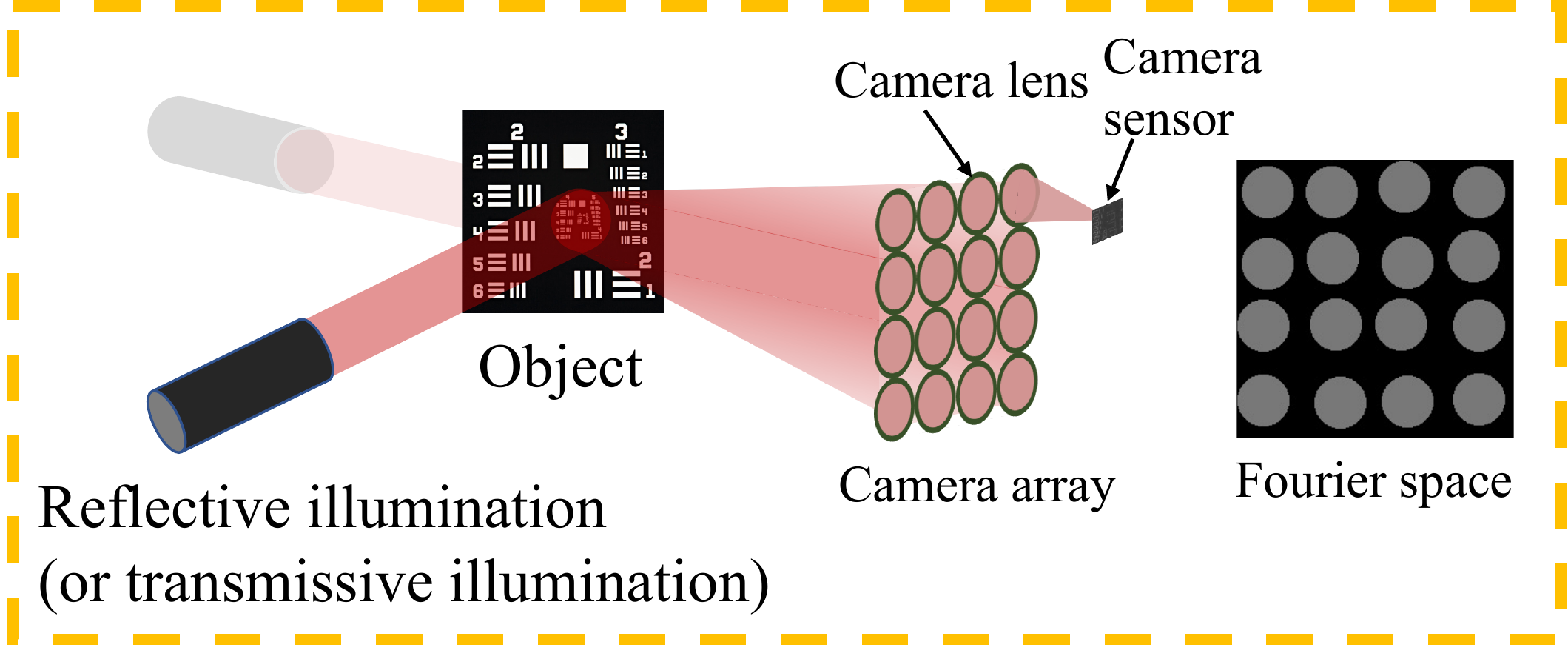}}
\hfill
\subfigure[Aperture-scanning FP~\cite{holloway2017savi}\label{fig:scheme_other}]{\includegraphics[width=0.491\textwidth]{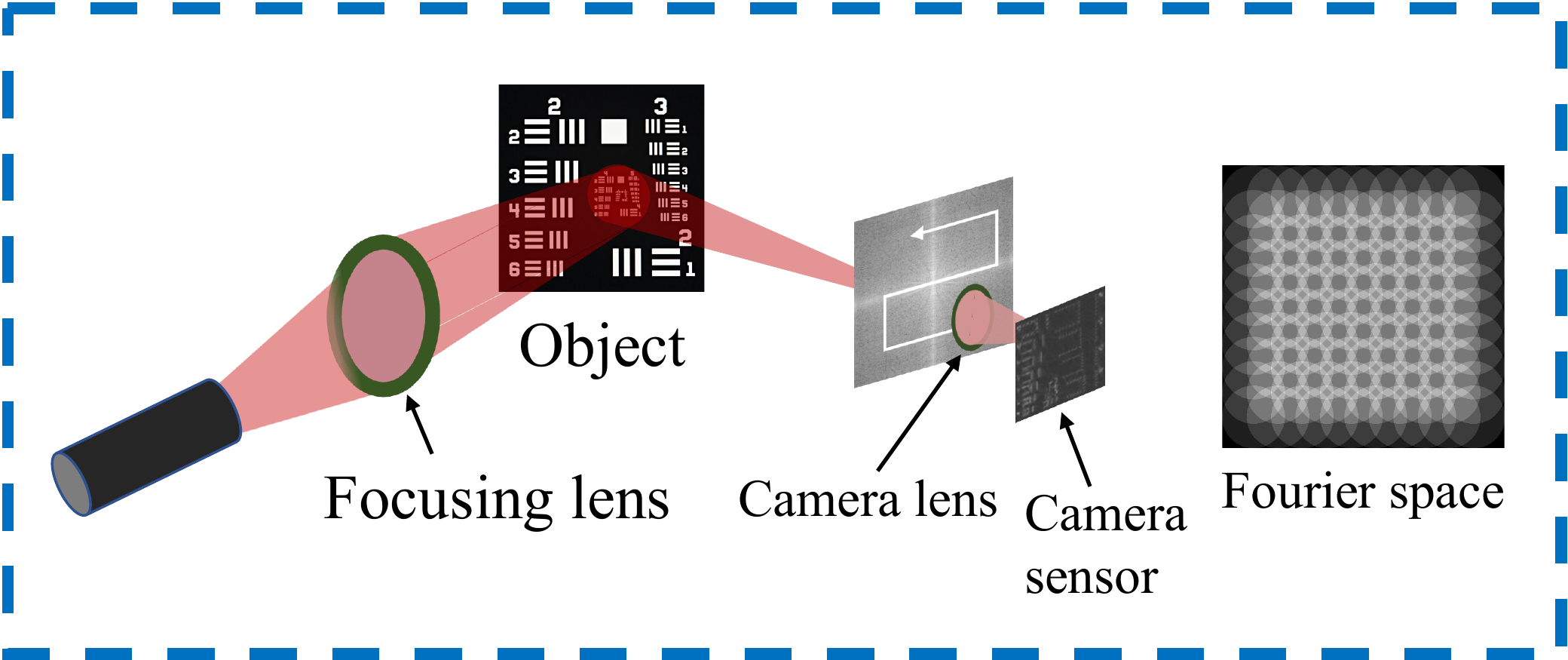}}
\caption{\label{fig:scheme}Comparison between the proposed method and conventional FP. (a)Proposed array camera snapshot FP. (b) Aperture-scanning FP~\cite{holloway2017savi}. The proposed method uses a sparse sampling strategy while conventional FP requires scanning the entire Fourier space in an overlapping manner.}
\end{figure}

We consider the array camera imaging system shown in Fig. \ref{fig:scheme_our}. The object is illuminated by coherent light source, such as laser, and captured by a camera array. One may roughly model each camera as a low pass filter on the object field with transfer function $H(u,v)=P(\lambda F u,\lambda F v)$, where $P(x,y)$ is the pupil function~\cite{brady2009optical} and $F$ is the focal length. In array of identical cameras, the transfer function for the $i^{th}$ camera centered at position $(x_i,y_i)$ is $P(\lambda F u-x_i,\lambda F v-y_i)$. The corresponding coherent impulse response for this camera is 
\begin{equation}
    \label{eq:psf}
    h_i(x,y)=e^{2\pi i\frac{x_ix+y_ix}{\lambda F}}h_o(x,y)
\end{equation}
where $h_o(x,y)$ is the point spread function for a camera at the center of the $(x,y)$ plane. In practice, camera tilt, focal state variation and uncertainty in $(x_i,y_i)$ impact how well the phase function $\phi_i(x,y)\approx 2\pi i\frac{x_ix+y_ix}{\lambda F}$ is known, but for present purposes it is sufficient to define the array measurement model as 
\begin{equation}
    \label{eq:forwardModel}
    g_i(x',y')=\left |\int\int f(x,y) e^{i\phi_i(x,y)}h_o(x'-x,y'-y) dx dy\right |^2
\end{equation}
Conventional Fourier ptychography uses iterative phase retrieval to invert the spectrogram given in Eqn.~\ref{eq:forwardModel}. Here, however, we propose to directly apply deep learning to estimate $f(x,y)$. This approach enables reconstruction from under sampled Fourier data and avoids the need to precisely characterize $\phi_i(x,y)$. 

In previous multiaperture FP studies, either the camera positions or the illumination wave direction is varied to enable oversampling of the target Fourier space. For example, Fig.~\ref{fig:scheme_other} shows the aperture-scanning FP~\cite{holloway2017savi}, where the camera moves to capture different regions of the Fourier space. The single frame Fourier coverage may be visualized by a disk in the object Fourier space, where the disk is defined by the pupil function. Shifting the phase $\phi_i$ by changing the camera position $x_i,y_i$ or by changing the coherent wave illumination angle shifts the position of the bandpass filter. Conventional FP assumes a dense array of overlapping bandpass measurements. Such sampling is not possible in a single frame of multicamera data. Rather we sparsely sample the Fourier space as illustrated at the right of Fig.~\ref{fig:scheme_our}.

\begin{figure}[htbp]
\centering
\includegraphics[width=0.9\textwidth]{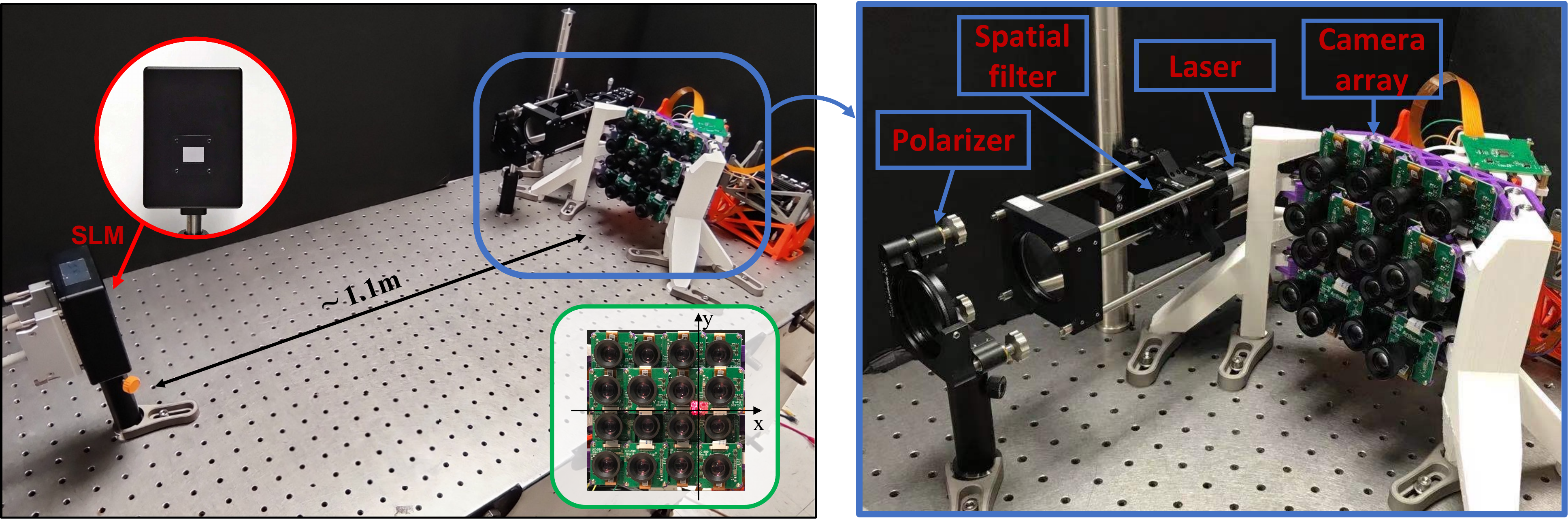}
\caption{\label{fig:proto}Experimental camera array and imaging system. }
\end{figure}

An experimental system built to validate the proposed approach is shown in Fig. \ref{fig:proto}. A superluminescent 650 nm light emitting diode (Exalos, Langhorne PA) was used for illumination. A spatial filter was utilized to collimate the source. Object patterns $f(x,y)$ were created using a liquid crystal spatial light modulator (SLM, Hamamatsu X10468), containing $600\times800$ pixels with pixel pitch 20$\mu m$. The reflected, phase modulated wave was imaged onto the camera array. All cameras in the array were focused on the SLM plane. The array consisted of 16 1MP OV9281 global shutter cameras (Arducam B0267) coupled with Marshall 25 mm f/2.5 lens (V-4325) operated on Nvidia Jetson Nano array. We 3D printed the supporting frame to mount the cameras in a $4\times4$ array. The sensors of the 16 cameras were not on the same plane, which allowed a slightly compact design. The offset between the optical axes of the adjacent lenses was approximately 33 mm. We also adjusted the orientation of each camera such that the target appeared at the center of its captured frame. 

To avoid grating diffraction from the pixelation of the SLM and maintain a proper measurement resolution, the SLM was placed 1.1 m away from the camera array. The period of the diffraction pattern at this distance was 35.7 mm. By letting the $0^{th}$ order diffraction fall into the gap between the right four cameras in the middle layers, no diffraction pattern was captured. At this distance, the target was measured by approximately $90\times120$ pixels on each camera.

The liquid crystal on silicon SLM modulates the phase in proportion to the voltage applied over each pixel~\cite{zhang2014fundamentals}. In our setup, the voltage is control by the pixel intensity, and higher intensity corresponds to lager phase retardation. Due to the pixel crosstalk caused by the fringing fields and elastic forces of the material, the expected phase retardation is spatially low pass filtered~\cite{persson2012reducing,zaperty2018numerical}. To mitigate this effect we limited our experiments to binary phase modulation with maximal retardation and clear high frequency images.

%The goal of this system is to jointly process array camera data to recover full resolution images from the down-sampled data. 
The pixel magnification was $600/90\approx6.7$, meaning that one camera pixel measures $6.7$ SLM pixels. The system goal is to jointly process the 16 array camera images to upsample to the original images. Such upsampling is possible because of the systematic variation in the subsampled images due to the phase functions $\phi_i(x,y)$. To demonstrate the feasibility of such upsampling, we used the physical array to measure output signals for several thousand images displayed on the SLM. We used the known input images as ground truth and the output image array as input to a convolutional neural network. We then trained the network to associate the ground truth images with the measured data.

\begin{figure}[htbp]
\centering
\includegraphics[width=0.60\textwidth]{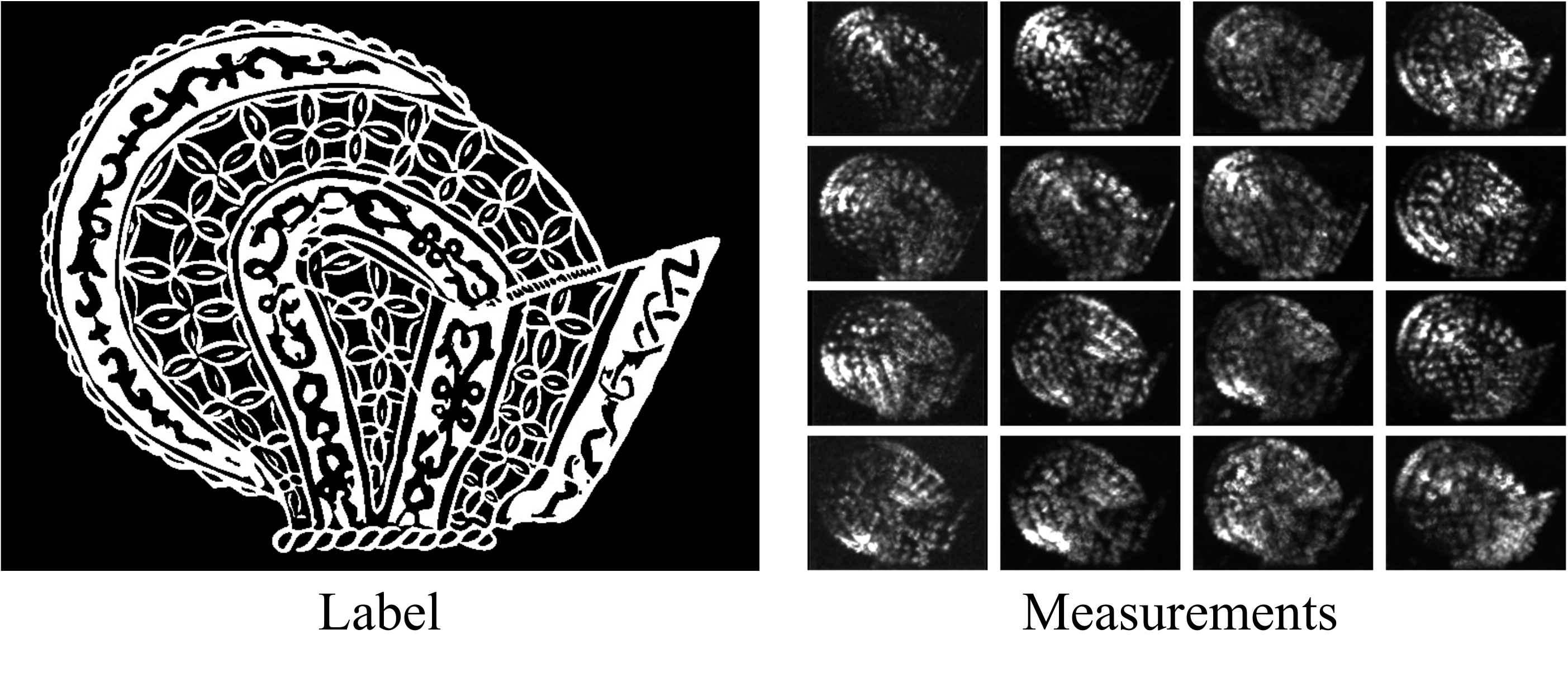}
% \hspace{1mm}
% \subfigure[Measurements]{\includegraphics[width=0.35\textwidth]{measurement_sample.pdf}}
\caption{\label{fig:measure_sample}A data sample on the SLM and the corresponding measurements from the camera array.}
\end{figure}

We first collected 2665 vector clip arts from Openclipart\footnote{\url{https://openclipart.org/}}. With image augmentation methods, i.e., rotation, flipping and scaling, we generated 23200 binary images with resolution $576\times768$ which were zero-padded to $600\times800$. The padding was applied because boundary pixels were occluded by the case of the SLM from some viewpoints. During the capture process, 5 frames were averaged for each camera to suppress noise. The low-resolution images from each camera were cropped to just the $90 \times 120$ region imaging the SLM. An example training image and its corresponding measurements are shown in Fig.~\ref{fig:measure_sample}.

\section{Image Estimation}
\label{sec:inverse}
\begin{figure}[htbp]
\centering\includegraphics[width=0.95\textwidth]{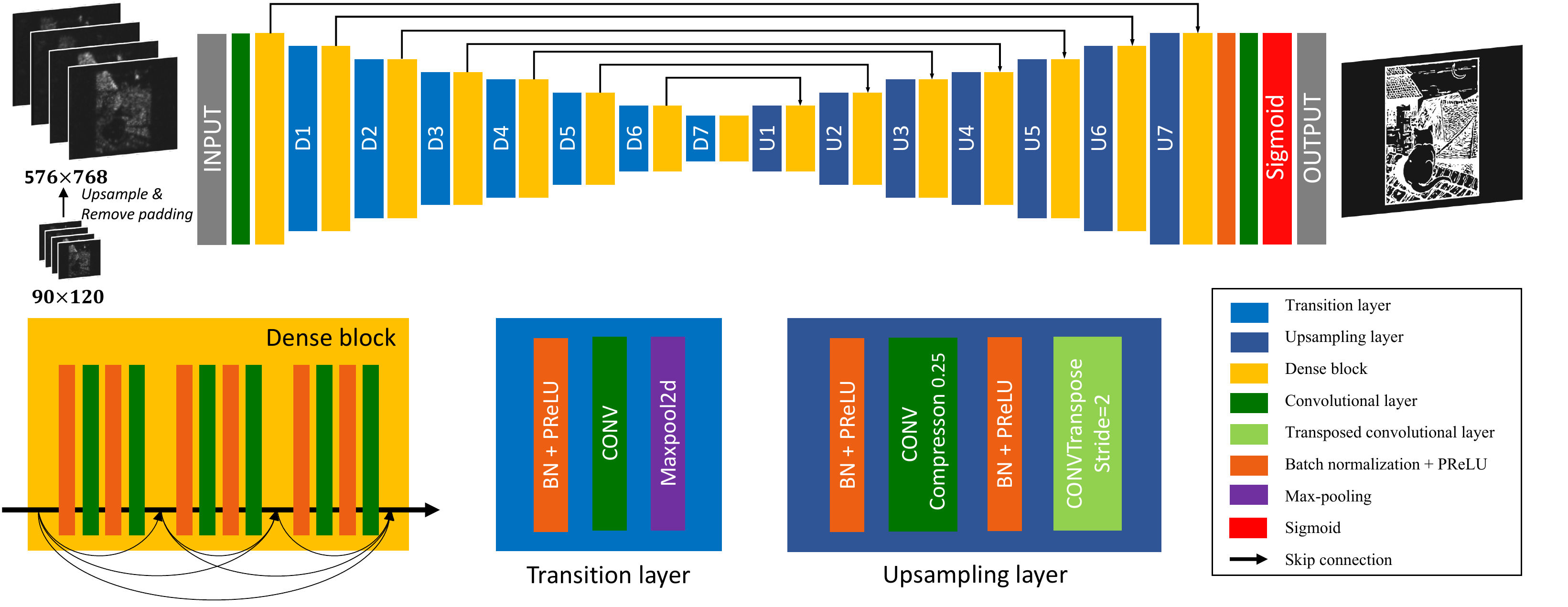}
\caption{\label{fig:proto_network}Illustration of the reconstruction network. The network adopts from the U-Net~\cite{ronneberger2015u} and the DenseNet~\cite{huang2017densely}.}
\end{figure}

The reconstruction network adopted the U-Net structure~\cite{ronneberger2015u} with dense blocks~\cite{huang2017densely}, as shown in Fig.~\ref{fig:proto_network}. The network consists of the initial convolutional layer, 14 dense blocks, 7 transition layers, 7 upsampling layers and a final convolutional layer. The definitions of the dense block and transition layer follow the original DenseNet~\cite{huang2017densely}. Each dense block consists of 5 BN-PReLU-Conv(1$\times$1)-BN-PReLU-Conv(3$\times$3) building blocks. The growth rate is $k = 24$ and each bottleneck layer produces $4k$ feature maps. The compression factor $\theta$ equals 0.8 in transition layers. The upsampling layer replaces the max-pooling in transition layer with deconvolution, and we apply compression factor $\theta = 0.2$. The initial convolutional layer produces 64 feature maps with filter size $3\times3$, and the final convolutional layer uses filter size $1\times1$. We upsampled the measurements so the input to the network matched the output image in spatial dimension.

Our 23200 image dataset was separated into 20000 training images and 3200 testing images. The network was first trained with binary cross-entropy loss, Adam optimizer~\cite{kingma2014adam} and learning rate 0.0003 for 100 epochs in PyTorch. After that we built a sub-dataset by selecting out the data with poor reconstruction SSIM. We tuned the network with the sub-dataset for 20 epochs using the following loss:
\begin{equation}
\label{eq:loss_function}
    l = l_{BCE} + \lambda l_{SSIM},
\end{equation}
where we selected $\lambda = 0.01$. This training method avoided the domination of smooth data~\cite{gharbi2016deep}. The training was performed on four Nvidia Tesla V100 with batch size 12. The resulting network, along with the code used to train it, is available for download at~\cite{code}.

We evaluated the network with widely used image quality assessment metrics, and the results are summarized in Table~\ref{tab:network_results}. We also show results for networks that used only subsets of the 16 measured images. (See the following text for details.) While the trend to improved image resolution is clear in the results, the effect of aperture synthesis is much clearer in actual images for the sparse binary patterns used here. 

\begin{table}[htbp]
\footnotesize
  \centering
  \caption{\rm Performance metrics evaluated on the testing data.}
    \begin{tabular}{|c|c|c|c|}
    \cline{1-4}
   Input&MSE & SSIM & BCE\\
    \cline{1-4}
      4 measurements  & 0.0561 & 0.7447 & 0.1767\\
    \cline{1-4}
      12 measurements & 0.0548 & 0.6974 & 0.1766\\
    \cline{1-4}
      16 measurements & 0.0428 & 0.8117 & 0.1776\\
    \cline{1-4}
    \end{tabular}%
  \label{tab:network_results}%
\end{table}%

Several reconstructed samples\footnote{In the current and the following sections, "reconstruction" refers to the network output after thresholding.} are demonstrated in Fig.~\ref{fig:slm_results_resolution}.  In each sample, the reconstructed image, the ground truth image, and the image directly down-sampled to $90\times120$ are shown. The down-sampled image represents the resolving power of a single camera in terms of the sensor pixel size. The images show that the proposed imaging method overcomes the pixel-limited resolution and super-resolves the texture details. It is worth-noting that the measurements did not include a bright field image as in traditional FP, so the low-frequency structural information was inferred from measurements. More samples along with their measurements are shown in Section 1 in Supplement 1.

\newcommand{\figwd}{0.163}
\newcommand{\drawfigure}[4]{
\tiny
    \includegraphics[width=\figwd\linewidth]{#1/output_thresh.png}
	\includegraphics[width=\figwd\linewidth]{#1/label.png}
	\includegraphics[width=\figwd\linewidth]{#1/downsample.png}
	\hspace{0.5mm}
	\includegraphics[width=\figwd\linewidth]{#2/output_thresh.png} 
	\includegraphics[width=\figwd\linewidth]{#2/label.png}
	\includegraphics[width=\figwd\linewidth]{#2/downsample.png}\\
	\vspace{0.3mm}
	\stackunder[2pt]{\includegraphics[width=\figwd\linewidth]{#3/output_thresh.png}}{Reconstruction} 
	\stackunder[2pt]{\includegraphics[width=\figwd\linewidth]{#3/label.png}}{Ground truth}  
	\stackunder[2pt]{\includegraphics[width=\figwd\linewidth]{#3/downsample.png}}{Down-sampled}  
	\hspace{0.5mm}
	\stackunder[2pt]{\includegraphics[width=\figwd\linewidth]{#4/output_thresh.png}}{Reconstruction} 
	\stackunder[2pt]{\includegraphics[width=\figwd\linewidth]{#4/label.png}}{Ground truth}  
	\stackunder[2pt]{\includegraphics[width=\figwd\linewidth]{#4/downsample.png}}{Down-sampled}\\
}
\begin{figure}[htbp]
    \centering
    \drawfigure{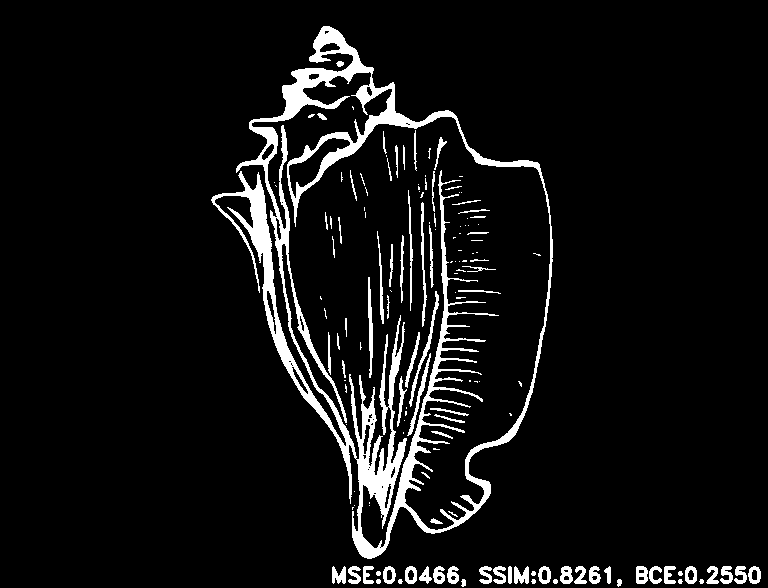}{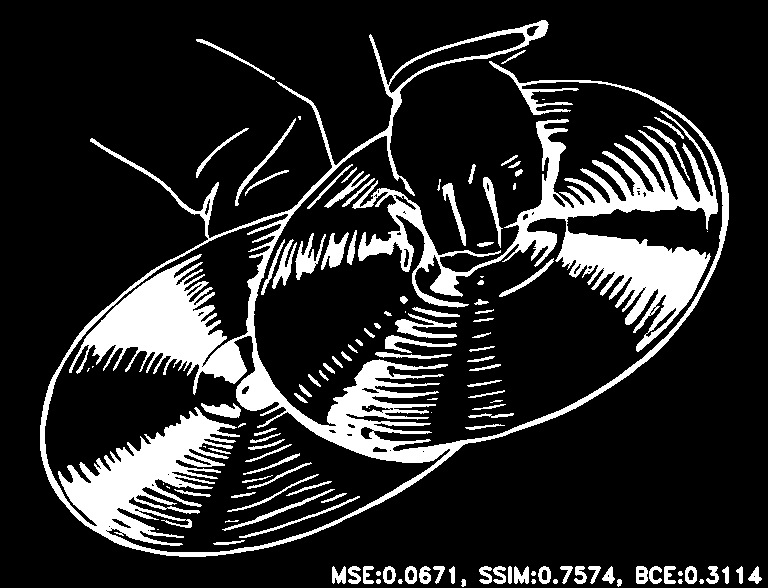}{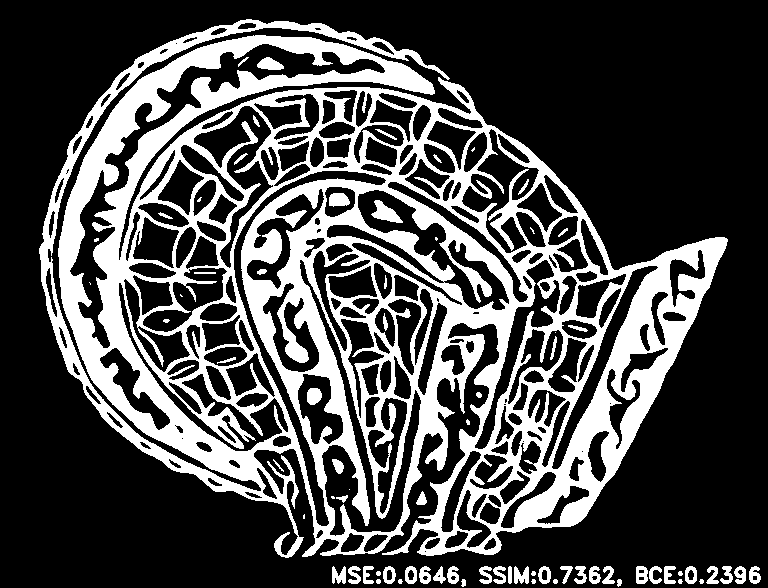}{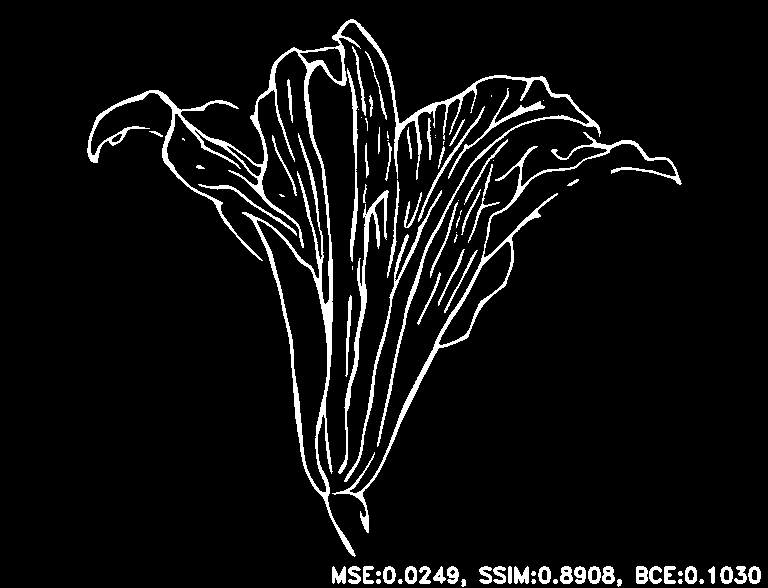}%{slm_data/good/20682}{slm_data/good/22028}
    \caption{\label{fig:slm_results_resolution} Improvement from pixel limited resolution. Zoom in to see details and quantitative evaluations}
\end{figure}

\begin{figure}[htbp]
\centering
\subfigure[\label{fig:slm_resolution_compare_full}Full images]{\includegraphics[width=0.215\textwidth]{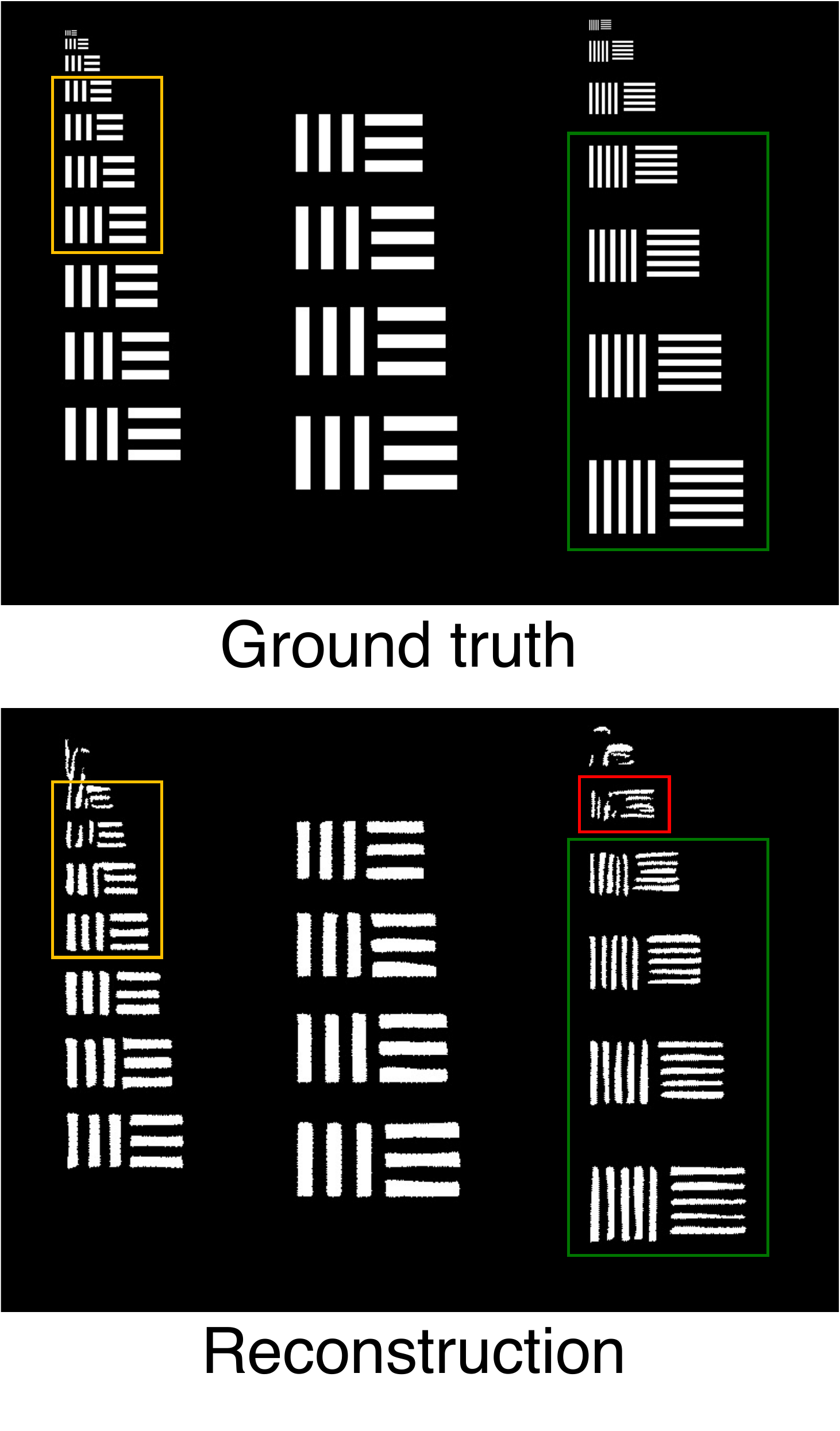}}
\hspace{0.05mm}
\subfigure[\label{fig:slm_resolution_compare_detail}Details of various bar groups]{\includegraphics[width=0.77\textwidth]{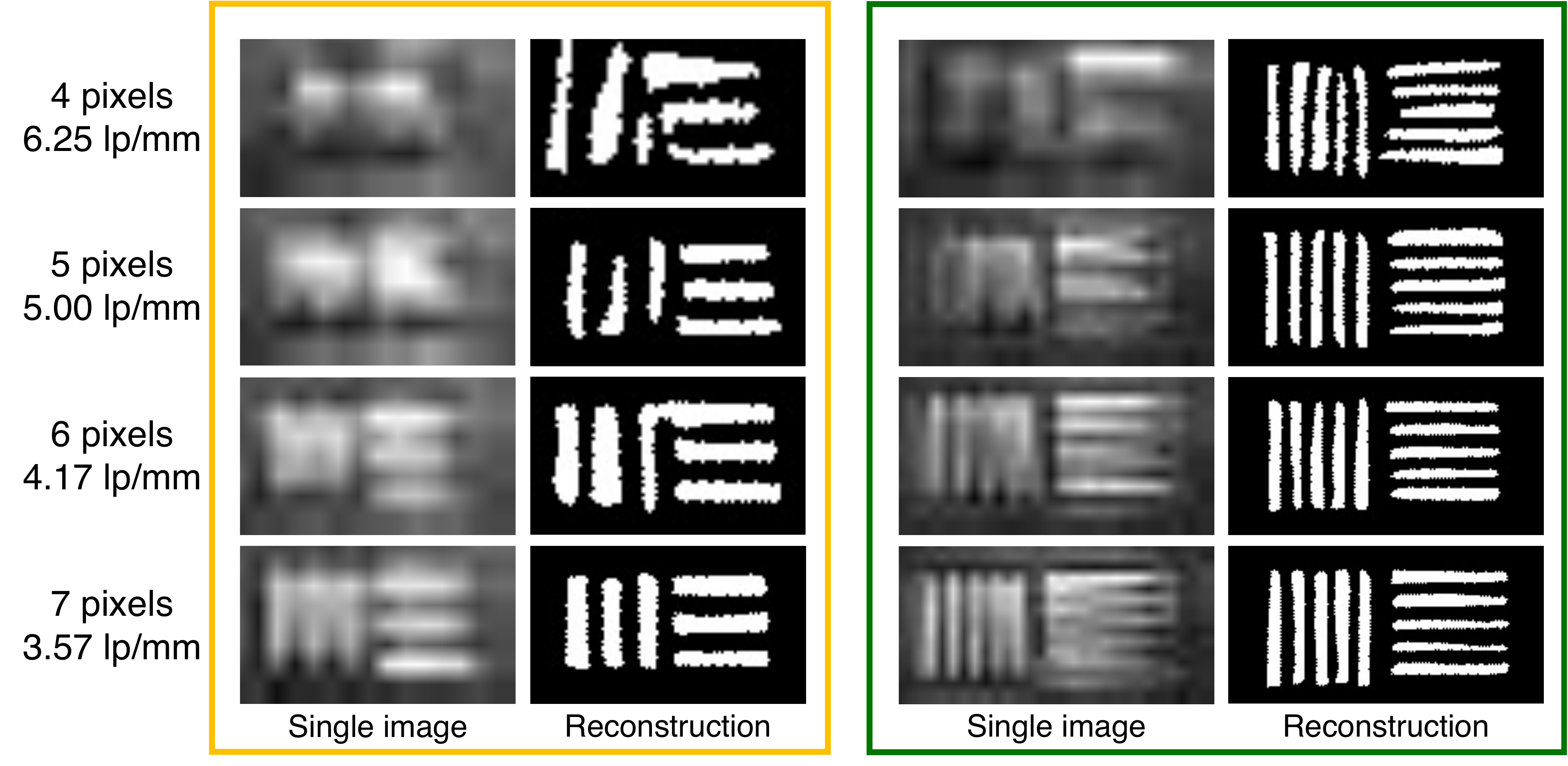}}
\caption{\label{fig:slm_resolution_compare} Resolution improvement from a single image. The bars with the width of 4 pixels can be resolved, and we can still see patterns when the width is 3 pixels (red bounding box in (a)). Zoom in to see details.}
\end{figure}

The resolution of a camera is limited by diffraction blur, geometric aberration and pixel sampling. We studied the actual resolution improvement of the system by imaging a resolution test chart. The target and the reconstructed image are shown in Fig.~\ref{fig:slm_resolution_compare_full}. The width of each line increases from 1 pixel to 14 pixels in the left two columns and from 1 pixel to 7 pixels in the right column. We also directly imaged the target with a single camera using a polarizer, and we compare the direct imaging and the reconstruction from 16 measurements in Fig.~\ref{fig:slm_resolution_compare_detail}. From direct imaging, the minimum resolvable width of a bar is 7 pixels on the SLM, which agrees with the down-sampling ratio of the camera. With the proposed imaging method, the width of a resolvable bar decreases to 4 pixels, and the we can still see repeating patterns in the right column when the width is 3 pixels.% diffraction blur:2.44*0.650*2.5

\begin{figure}[htbp]
\tiny
\centering
\includegraphics[width=\figwd\linewidth]{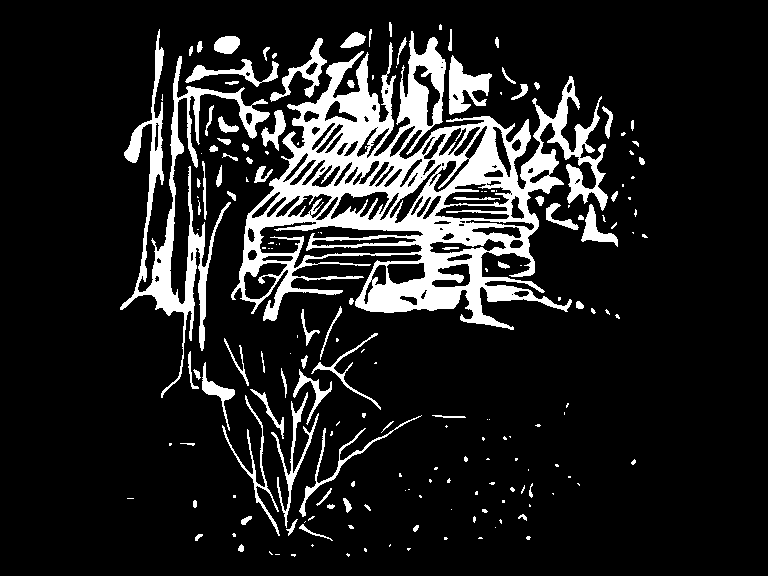}  
\includegraphics[width=\figwd\linewidth]{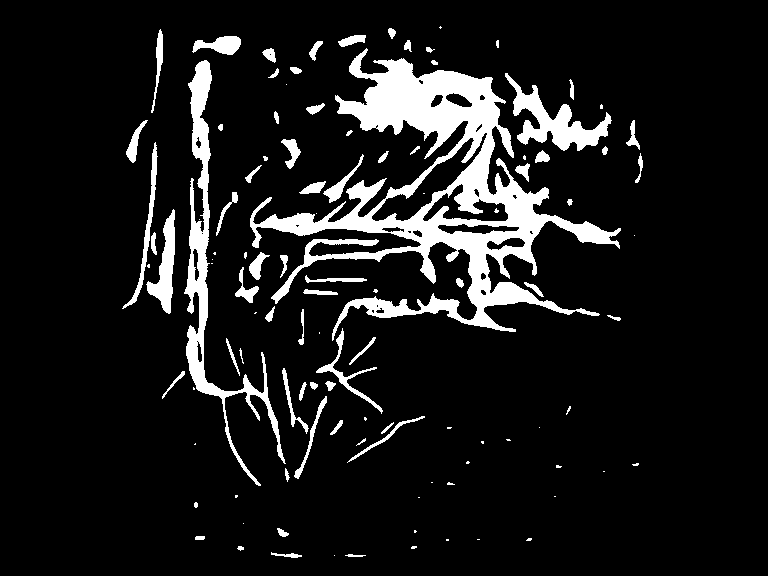}  
\includegraphics[width=\figwd\linewidth]{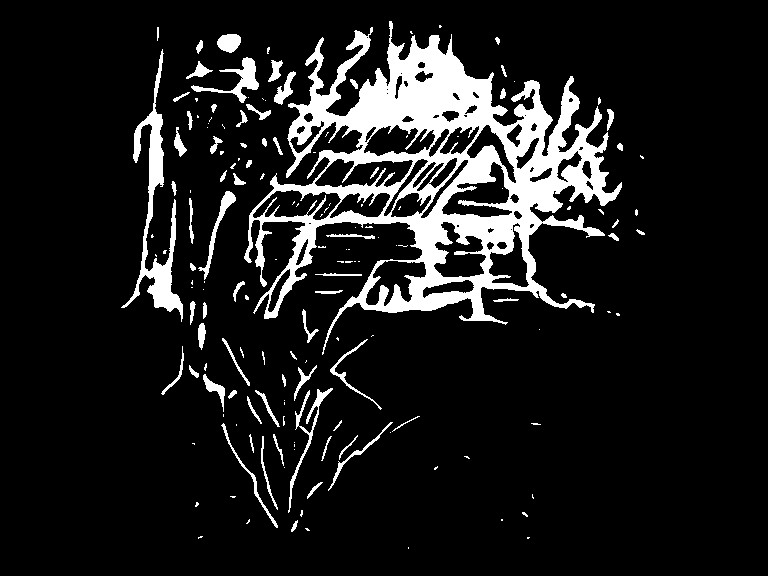}  
\includegraphics[width=\figwd\linewidth]{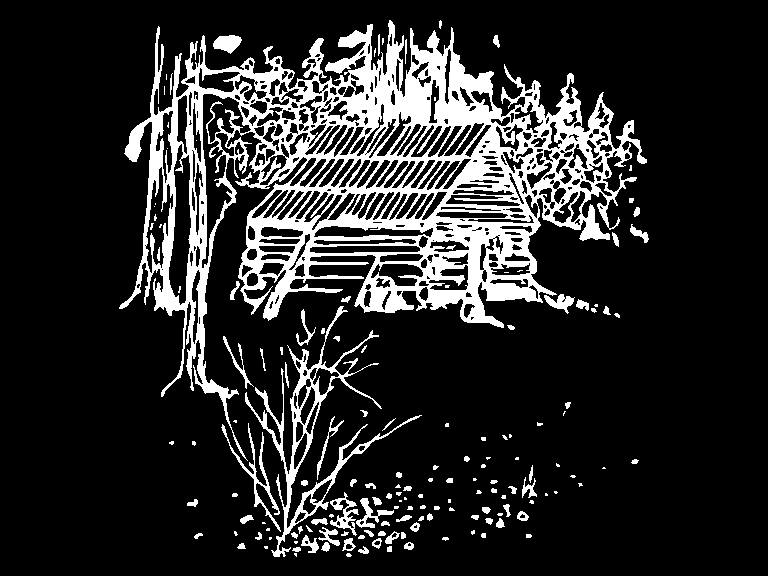} \\ 
\includegraphics[width=\figwd\linewidth]{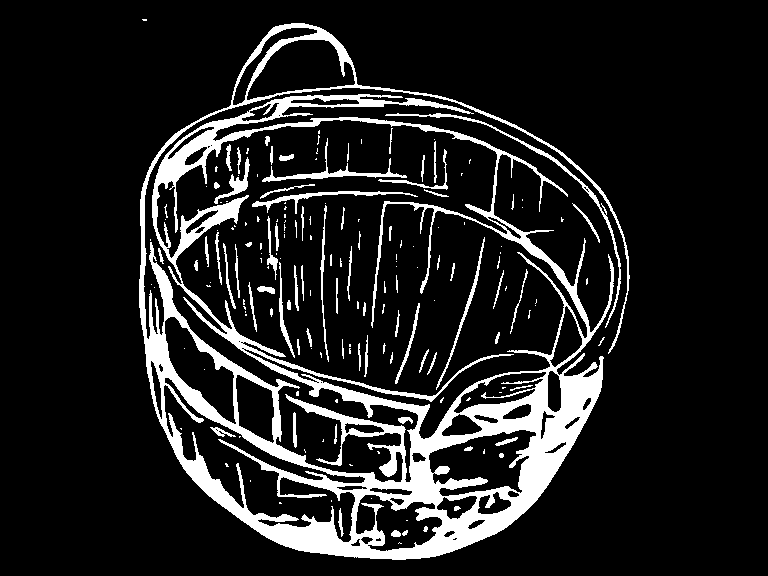}  
\includegraphics[width=\figwd\linewidth]{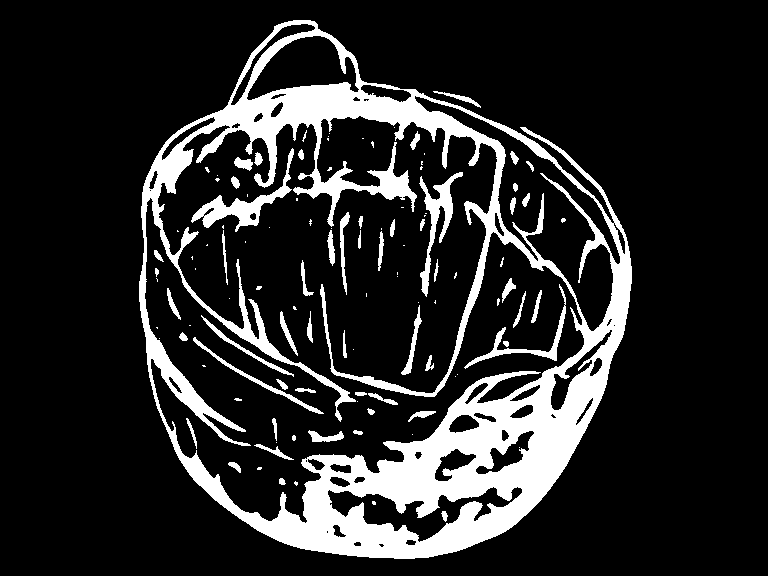}  
\includegraphics[width=\figwd\linewidth]{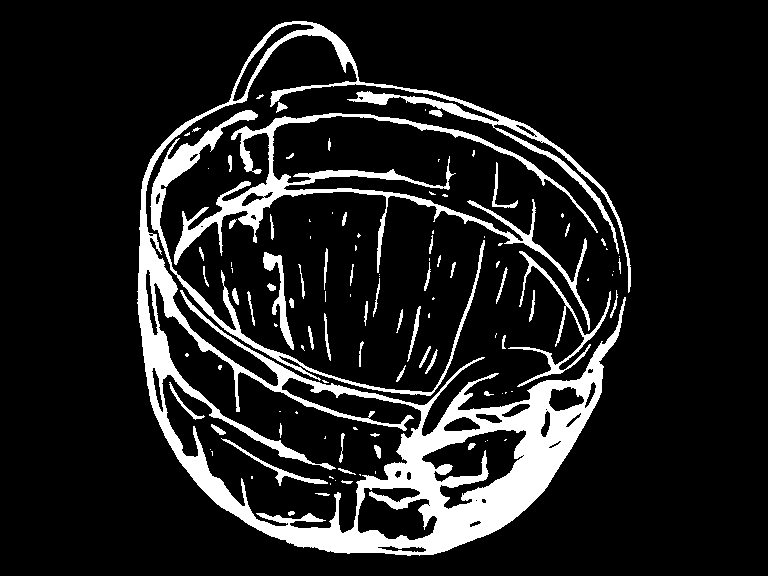}  
\includegraphics[width=\figwd\linewidth]{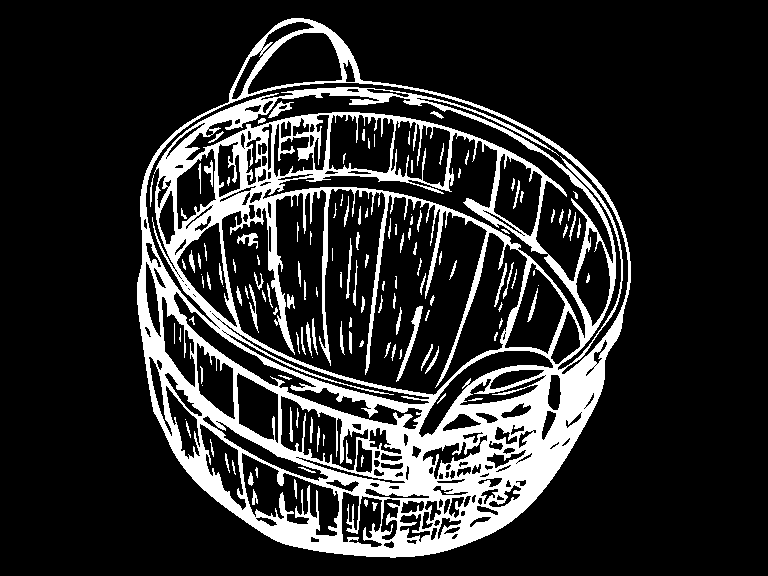} \\ 
\stackunder[2pt]{\includegraphics[width=\figwd\linewidth]{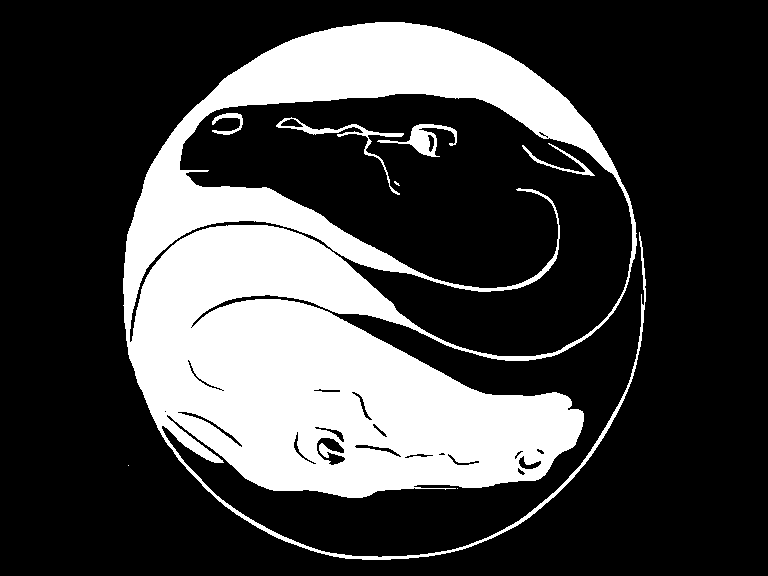}}{16 Measurements}  
\stackunder[2pt]{\includegraphics[width=\figwd\linewidth]{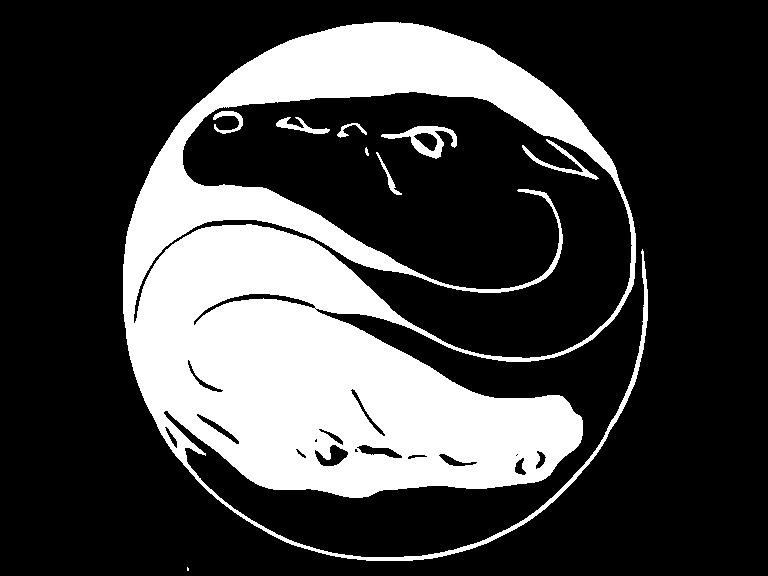}}{4 Measurements}   
\stackunder[2pt]{\includegraphics[width=\figwd\linewidth]{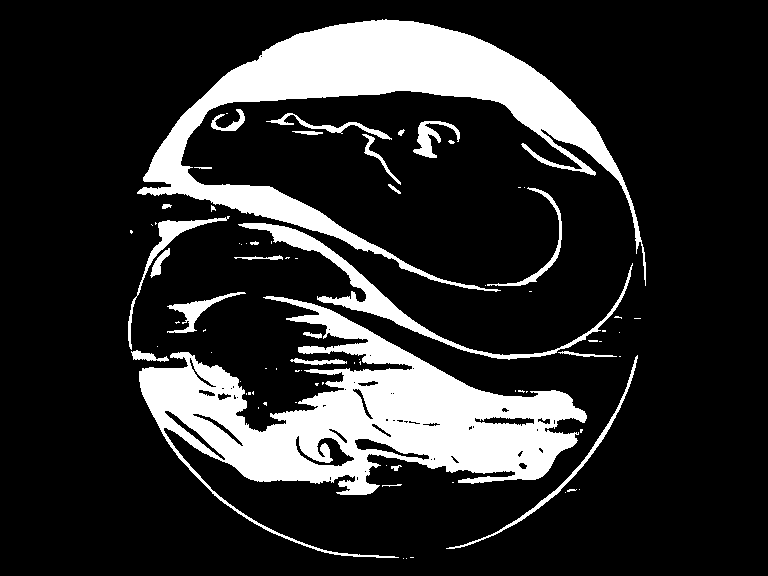}}{12 Measurements}   
\stackunder[2pt]{\includegraphics[width=\figwd\linewidth]{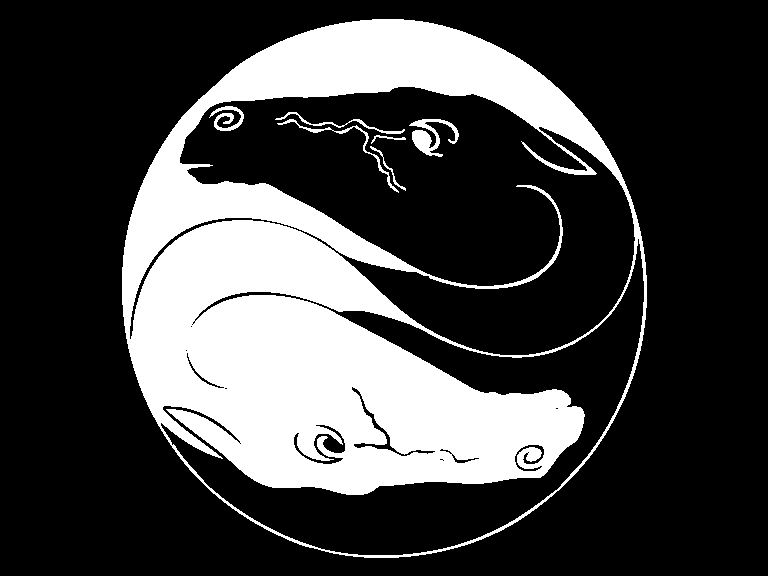}}{Ground truth}  \\ 
\caption{\label{fig:compare_differet_measurement}Comparing the reconstruction results using different number of measurements.}
\end{figure}

To confirm that our reconstructed image quality is based on aperture synthesis over the full camera array we trained two more networks that used only subsets of the 16 measured images. The first network used the images from the right four cameras in the middle layers, and the second network used the remaining 12 images. The quantitative results are shown in Table~\ref{tab:network_results}, and visual results are shown in Fig.~\ref{fig:compare_differet_measurement}. As one would expect, measurements close to the optical axis contribute to the reconstruction of low-frequency information, and the system relies on off-axis measurements to recover high-frequency details.

\newcommand{\drawpoorfigure}[4]{
\tiny
    \includegraphics[width=\figwd\linewidth]{#1/output.png}  
    \includegraphics[width=\figwd\linewidth]{#1/output_thresh.png}
	\includegraphics[width=\figwd\linewidth]{#1/label.png}
	\hspace{0.5mm}
	\includegraphics[width=\figwd\linewidth]{#2/output.png} 
	\includegraphics[width=\figwd\linewidth]{#2/output_thresh.png} 
	\includegraphics[width=\figwd\linewidth]{#2/label.png}\\
	\vspace{0.3mm}
	\stackunder[2pt]{\includegraphics[width=\figwd\linewidth]{#3/output.png}}{Network output} 
	\stackunder[2pt]{\includegraphics[width=\figwd\linewidth]{#3/output_thresh.png}}{Reconstruction} 
	\stackunder[2pt]{\includegraphics[width=\figwd\linewidth]{#3/label.png}}{Ground truth} 
	\hspace{0.5mm}
	\stackunder[2pt]{\includegraphics[width=\figwd\linewidth]{#4/output.png}}{Network output} 
	\stackunder[2pt]{\includegraphics[width=\figwd\linewidth]{#4/output_thresh.png}}{Reconstruction} 
	\stackunder[2pt]{\includegraphics[width=\figwd\linewidth]{#4/label.png}}{Ground truth}\\
}
\begin{figure}[htbp]
    \centering
    \drawpoorfigure{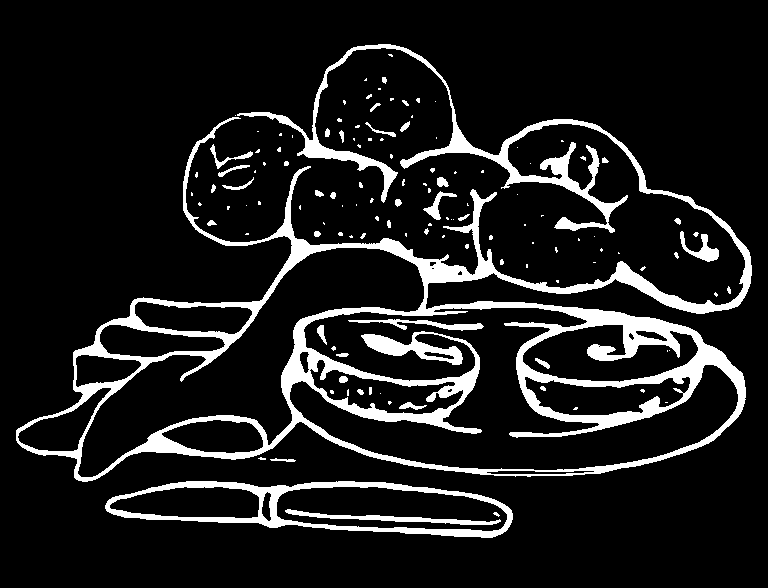}{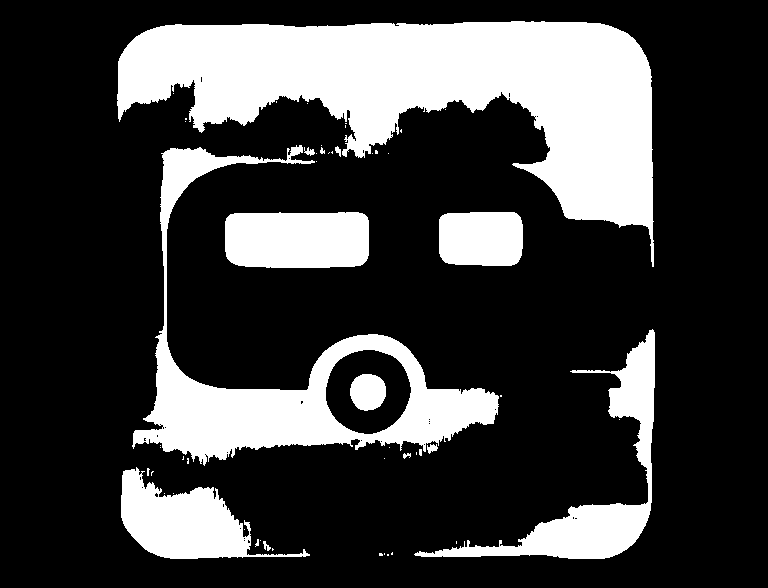}{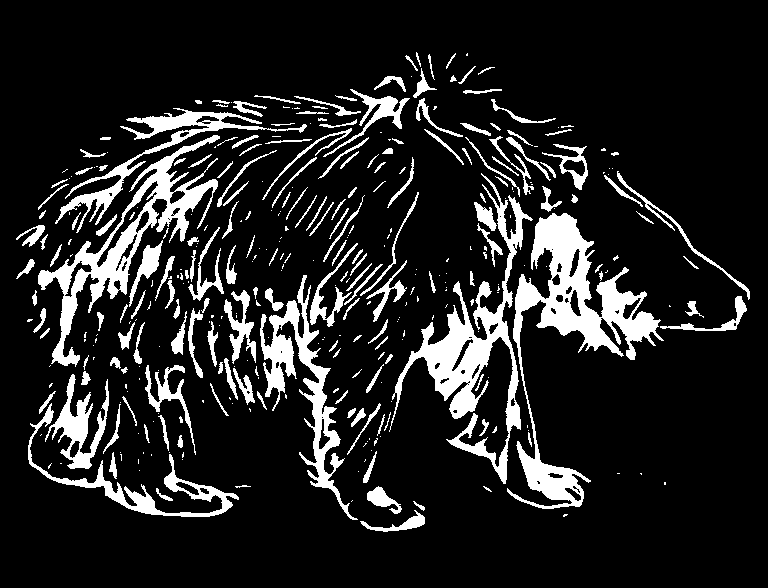}{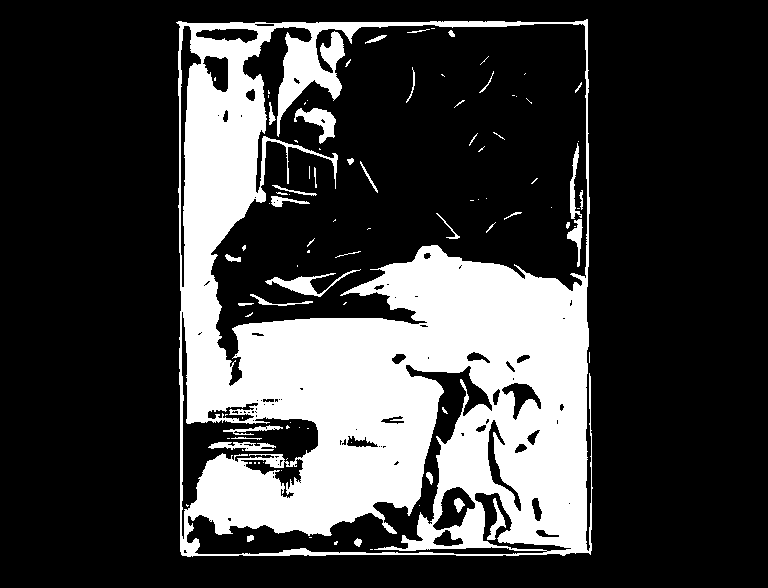}
    \caption{\label{fig:slm_results_poor}Representative samples with reconstruction errors.}
\end{figure}

Fig.~\ref{fig:slm_results_poor} shows example images with less satisfactory reconstruction results, which also represent the typical types of error in the testing data. The most common error comes from the random dots or large areas without phase variation in the image. Ptychographic image synthesis is based on self-referencing interference between adjacent pixels. For discrete point sources and other sparse images, such interference does not occur and aperture synthesis is impossible. The second type of error comes from animal images which consist of plenty of short lines with varying orientations. While lines are easier to reconstruct compared with dots, the varying random orientation still poses challenges in generating a binary image, but we are able to observe the texture from the network output before thresholding. We also see significant error in artistic images which is difficult to avoid due to the lack of similar samples in the training set. To improve the reconstruction fidelity and build systems for wider applications, the following aspects should be considered.

\textbf{Calibration}. One challenge in traditional FP is calibration, because the reconstruction algorithm requires an accurate forward model. In our experiments, we manually cropped the SLM region from the image without careful pixel alignment, and we did not characterize the pixel cross-talk on SLM. While the calibration can be implicitly completed by the neural network, the burden can be lifted to improve the network's resolving power. 

\textbf{Camera arrangement}. In our experiments, we did not directly measure the pattern on the SLM, and the results show that low-frequency information can be inferred from measured high-frequency information. However, adding a direct measurement should improve the reconstruction, especially when the image consists of mainly low-frequency components. In Sec.~\ref{sec:simulations}, we further discuss other considerations in the aperture distribution.

\begin{figure}[htbp]
\centering
\subfigure[\label{fig:geo_sample}A training sample with random geometrical shapes]{\includegraphics[width=0.24\textwidth]{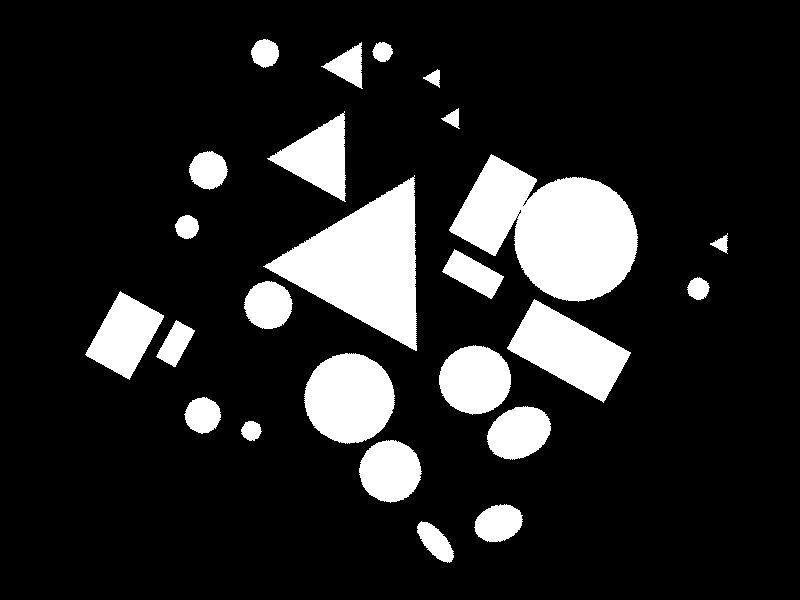}}
\hspace{0.05mm}
\subfigure[\label{fig:geo_gt}Sample ground truth]{\includegraphics[width=0.24\textwidth]{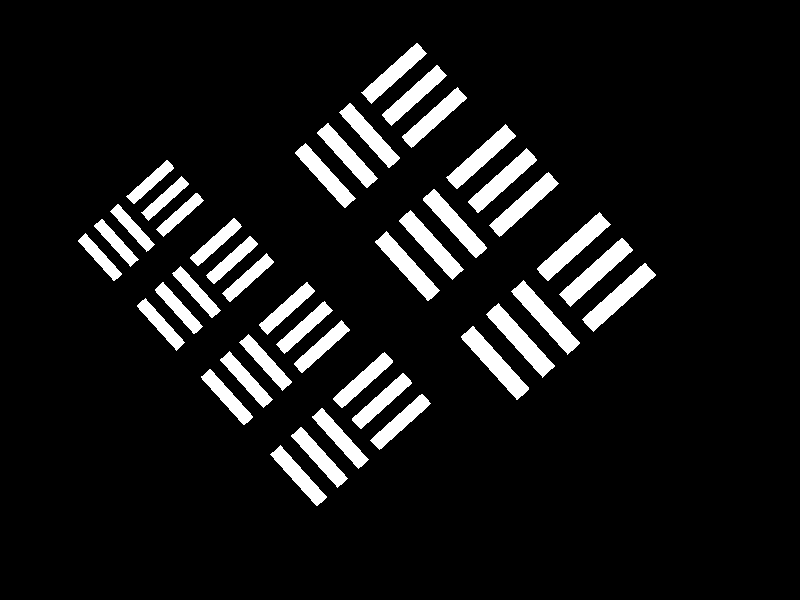}}
\hspace{0.05mm}
\subfigure[\label{fig:geo_with_geo}Reconstruction]{\includegraphics[width=0.24\textwidth]{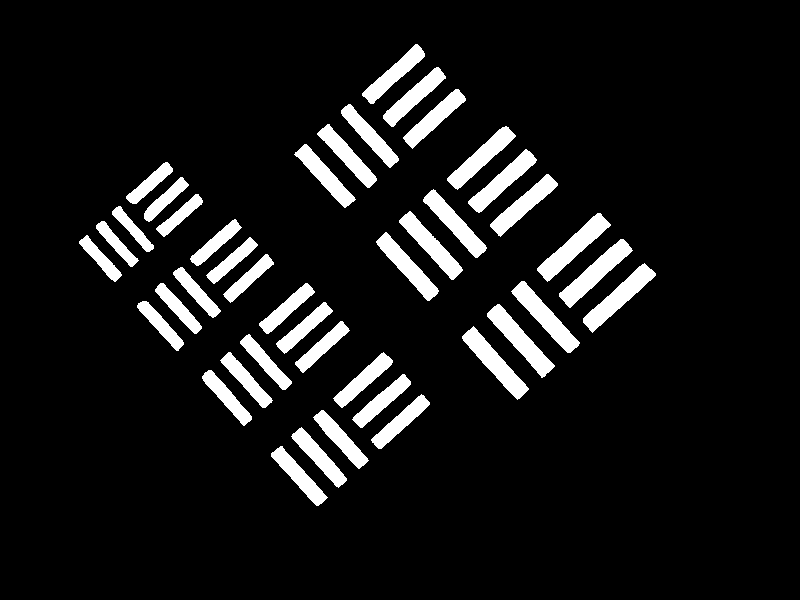}}
\hspace{0.05mm}
\subfigure[\label{fig:geo_without_geo}Reconstruction without geometrical training data]{\includegraphics[width=0.24\textwidth]{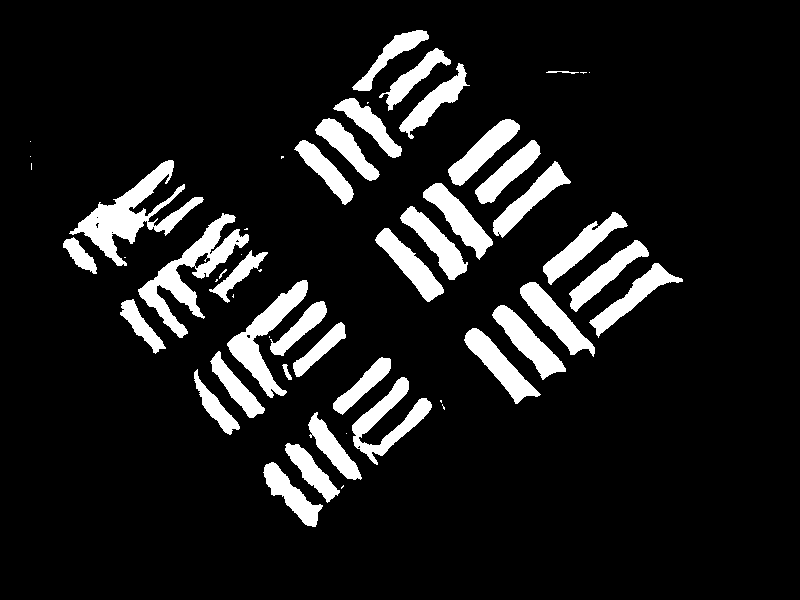}}
\caption{Effect of training data on reconstruction performance. (a) Samples with geometrical shapes were included in the training set. (b) A sample in testing data. (c) The testing sample could be exactly reconstructed. (d) Reconstruction showed significant artifacts when the geometrical shapes were removed from training data.}
\end{figure}

\begin{figure}[htbp]
\centering
\subfigure[Before tuning]{\includegraphics[width=0.24\textwidth]{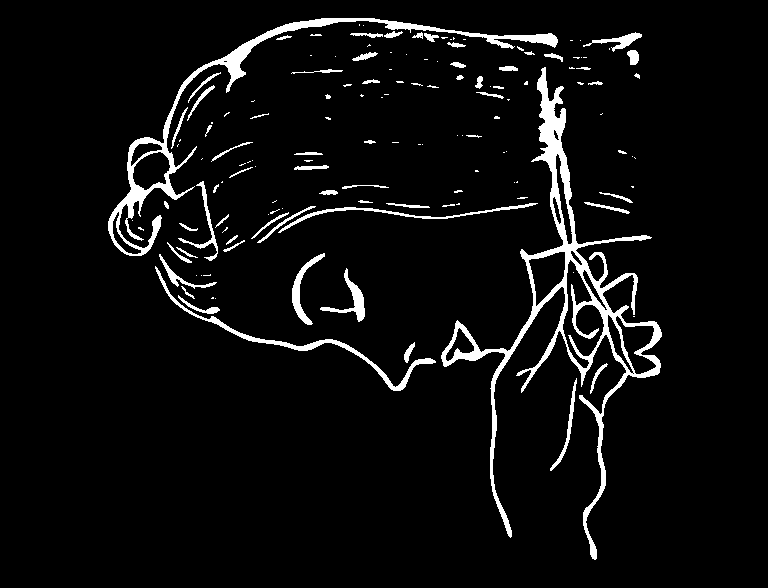}}
\hspace{0.05mm}
\subfigure[After tuning]{\includegraphics[width=0.24\textwidth]{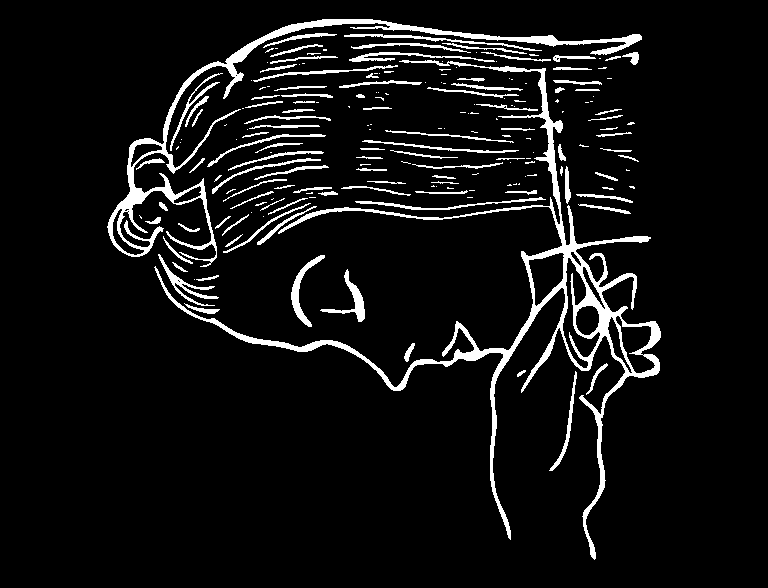}}
\hspace{0.05mm}
\subfigure[Ground truth]{\includegraphics[width=0.24\textwidth]{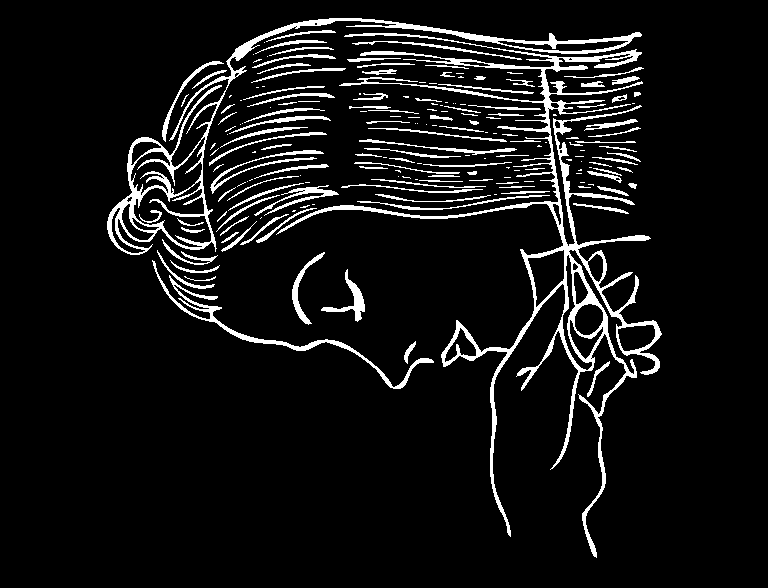}}
\caption{\label{fig:compare_trick}Effect of tuning with challenging sub-dataset.}
\end{figure}

\textbf{Dataset and training}. The performance of a neural network highly depends on the training data. In our experiments, the training data included 1880 samples containing random geometrical shapes, as shown in Fig.~\ref{fig:geo_sample}, and the network could reconstruct Fig.~\ref{fig:geo_gt} without perceptible error, see Fig.~\ref{fig:geo_with_geo}. In contrast, the reconstructed image, Fig.~\ref{fig:geo_without_geo}, showed obvious artifacts when the geometrical data were removed from the training set. The performance of the network is also affected by the training strategy. Fig.~\ref{fig:compare_trick} compares the reconstruction performance before and after the network was tuned with challenging sub-dataset. The reconstruction on details improved with this training trick.

\section{Design Analysis}
\label{sec:simulations}
In this section, we simulate diverse systems to consider how the experimental results presented above might improved. While the proposed method does not require an accurate forward model for reconstruction, the reconstruction fidelity is naturally affected by the diameter(s) and the distribution of the camera apertures. Here we simulate 10 aperture distribution strategies and compare their performances.

The simulation follows the traditional FPM as described in \cite{zheng2016fourier}, and for simplicity we drop the phase factor and the coordinate scaling. The complex wave from the object is denoted $\psi(x,y)$, and the field at the Fourier plane is denoted $\hat{\psi}(x',y')$, then the image measurement by the $i^{th}$ camera can be expressed as
\begin{equation}
\label{eq:noise_free_measurment}
    I_i(x,y) = |\mathcal{F}[\hat{\psi}(x',y')A(x'-x'_i,y'-y'_i)]|^2
\end{equation}
where $\mathcal{F}$ is the Fourier transform and $A(x'-x'_i,y'-y'_i)$ is the aperture centered at $(x'_i,y'_i)$ defined as
\begin{equation}
% A(x',y')=\left\{
% \begin{aligned}
% 1  &{x'^2+y'^2\leq(\frac{d}{2})^2}\\
% 0  &{otherwise}\\
% \end{aligned}
% \right.
  A(x',y')=
  \begin{dcases*} 
  1, & $x'^2+y'^2\leq(\frac{d}{2})^2$ \\ 
  0,  &{otherwise}
  \end{dcases*} 
,
\end{equation}
where $d$ is the diameter of the aperture. In simulations, we assumed the wave from the object $\psi(x,y)$ was a real-valued image with $512\times512$ pixels. 

In the first 4 strategies, we considered 9 apertures with aperture diameter $d=128$. 
Strategy 1 assumed an intuitively ideal but physically challenging layout that 9 apertures were densely located at the center of the Fourier space. 
Strategy 2 assumed an uniform, symmetrical and sparse distribution. 
Strategy 3 assumed an uniform, asymmetrical and sparse distribution. 
Strategy 4 assumed sparse but loosely structured distribution where each aperture was given a \textbf{random} shift compared to the strategy 2, and this strategy best described our physical setup.

\begin{figure}[htbp]
\centering\includegraphics[width=0.99\textwidth]{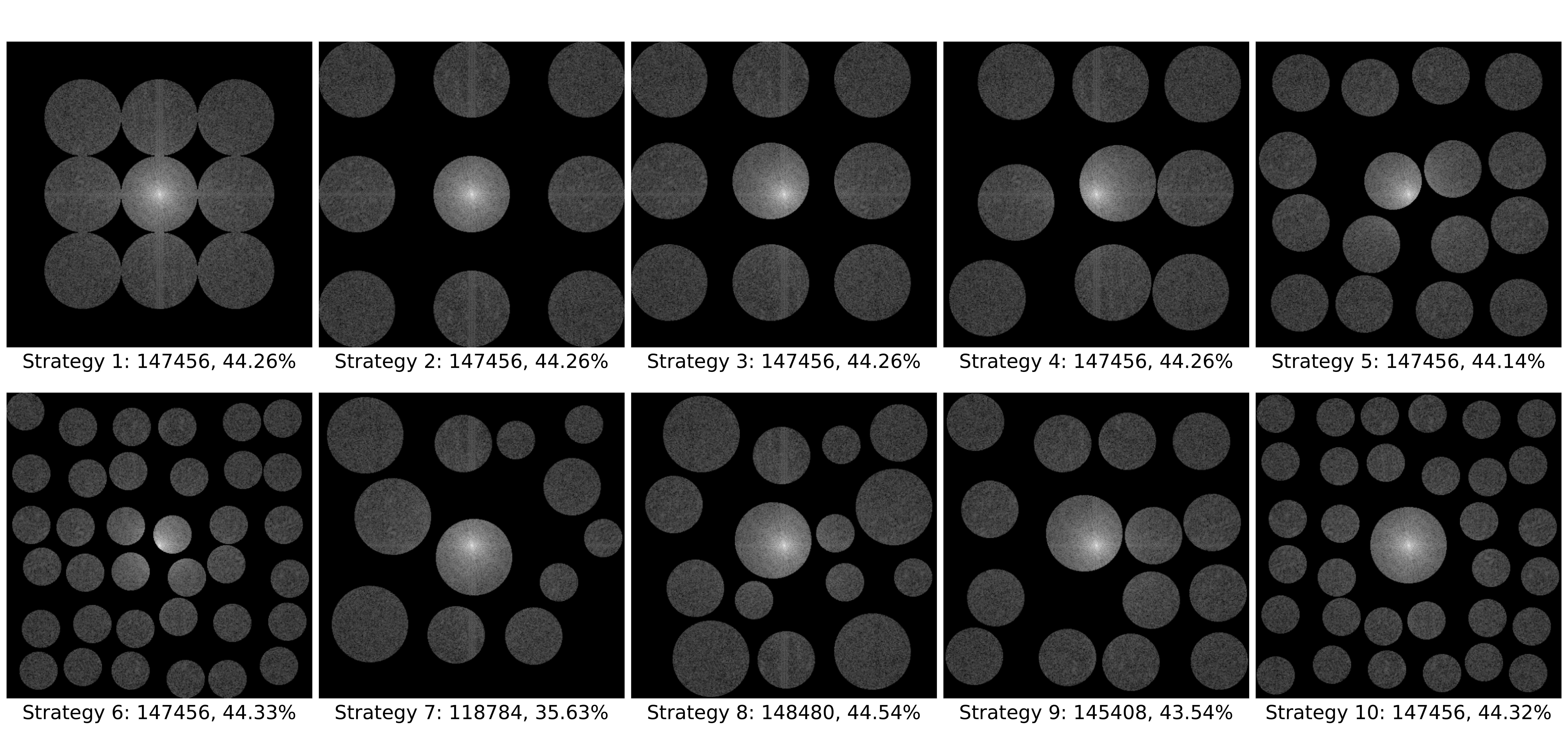}
\caption{\label{fig:distribution}Different aperture distribution strategies. Strategy 1-6 consider single aperture diameter, and Strategy 7-10 consider multi-scale aperture diameters. Strategy 7 consists of $d = 128, 96$ and 64, each with 4 apertures. Strategy 8 consists of $d = 128, 96$ and 64, each with 5 apertures. Strategy 9 consists of $d = 128$ and 96, with 1 and 14 apertures respectively. Strategy 10 consists of $d = 128$ and 64, with 1 and 32 apertures respectively. The total measured pixels and the measured percentage of the Fourier space are labeled under each strategy.}
\end{figure}

Strategy 5 and 6 considered 16 and 36 apertures with diameters $d = 96$ and $d = 64$ respectively.
Strategy 7-10 considered \textbf{multi-scale} aperture diameters and random distribution. The number of apertures and diameters for each distribution are illustrated in Fig.~\ref{fig:distribution}. The number of apertures were selected so that all strategies except strategy 7 had similar total measured pixels, while strategy 7 measured approximately 20\% less pixels.

The simulation data were generated from DIV2K dataset \cite{Agustsson_2017_CVPR_Workshops} and CLIC dataset~\cite{clic}. We cropped 20000 patches, each containing $512\times512$ pixels. The 20000 data were divided into 15000 training data, 2500 validation data and 2500 testing data. We considered the Poisson noise by introducing a parameter $n$ representing the expected number of photons, so the target was $\psi(x,y) = I_{gt}(x,y)$, where $ I_{gt}$ represented the image normalized to $[0,1]$, and the measurement became:
\begin{equation}
    I_i(x,y) = Poisson(|\mathcal{F}[\hat{\psi}(x',y')A(x'-x'_i,y'-y'_i)]|^2\times n).
\end{equation}

The reconstruction still adopted the U-net structure as shown in Fig.~\ref{fig:proto_network} but with 6 transition layers and 6 upsampling layers. The growth rate was 16, and each dense block consisted of 5 building blocks. We also considered the residual learning scheme~\cite{zhang2017beyond} and asked the network to predict the residual of the bright field low resolution image.

We first trained the network on noise-free data following Eq.~\ref{eq:noise_free_measurment}, and we tuned the network with noisy data with ($n=10^3$). The performance of the networks on the testing data with different noise levels are summarized in Table \ref{tab:simulation_compare_sparse}, and three reconstruction samples are shown in Fig.~\ref{fig:simulation_results}. Extra samples and full resolutions images are shown in Section 2 in Supplement 1. We emphasize the following observations:

\begin{enumerate}
\item \textbf{Sparsity}: Although the dense distribution better preserved the structure information of the image, its ability to resolve high frequency information is limited. In contrast, sparse measurement strategies captured more high-frequency information and recovered more details while stilling maintaining high PSNR. Dense distribution also showed poor robustness to noise compared to sparse distributions.

\item \textbf{Random distribution}: Given the diameter and the number of the apertures, as shown in strategy 2-4, randomly distributed apertures outperformed others in both quantitative evaluation and visual results.

\item \textbf{Multi-scale apertures}: While the resolving power of the system decreased with the aperture diameter in single-aperture-size cases, improved results were demonstrated by combining multi-scale apertures. This strategy even achieved competitive results with fewer measured pixels.
\end{enumerate}

\begin{figure}[htbp]
\centering
\includegraphics[width=0.99\textwidth]{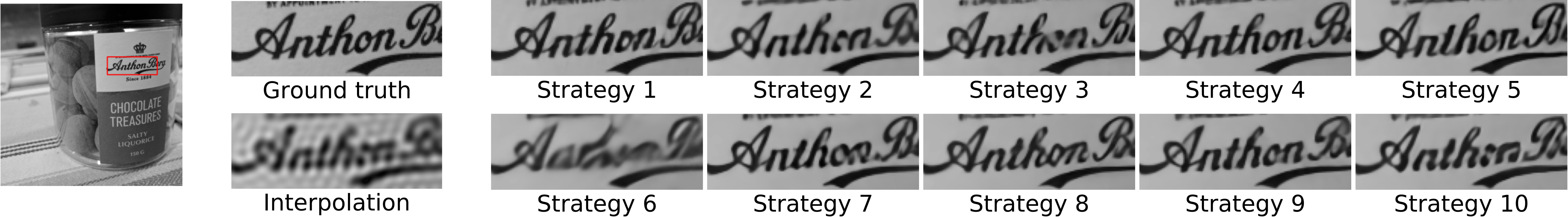}\\
\vspace{2mm}
\includegraphics[width=0.99\textwidth]{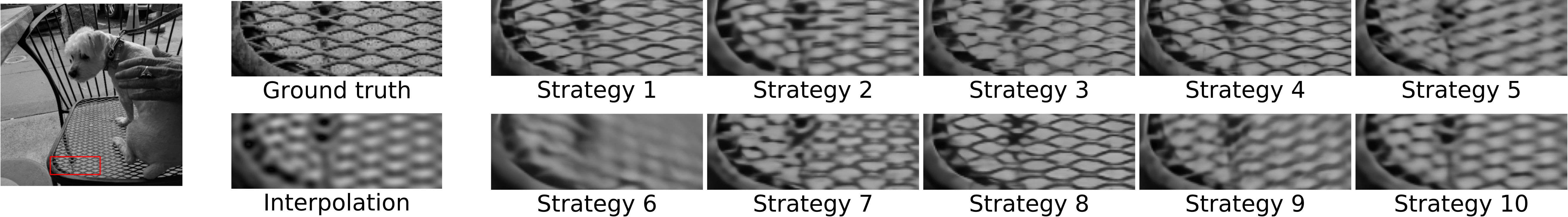}\\
\vspace{2mm}
\includegraphics[width=0.99\textwidth]{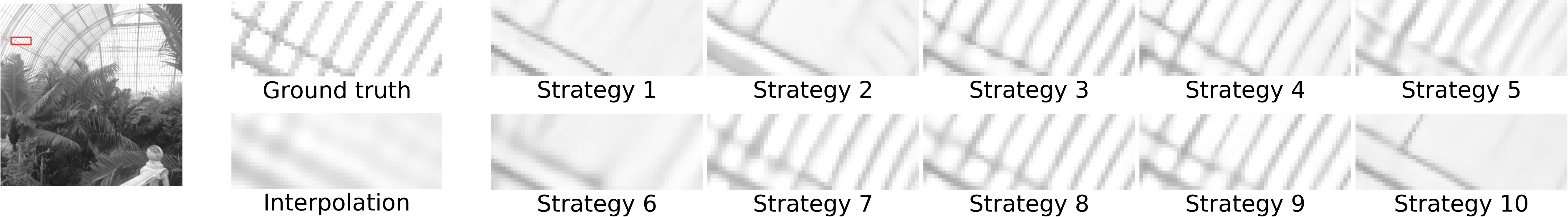}
\caption{\label{fig:simulation_results}Comparing the reconstruction results with different aperture distributions.}
\end{figure}

The first two observations allow great flexibility in camera array design, and the third observation allows using smaller apertures without compromising the resolving power of the system. These pave the way for developing cheap and portable system for wider applications. 

While we applied the same network structure to all strategies to ensure fair comparison, we admit that improved results could be achieved with advanced network structures. In fact the state-of-the-art single image super-resolution algorithms easily exceed 10 million parameters~\cite{Guo_2020_CVPR}, and we had only 5 million parameters in contrast. The network structure also ignores the different scales of the measurements. Future research can focus on jointly optimizing the aperture distribution and the reconstruction network.

\begin{table}[htbp]
\footnotesize
  \centering
  \caption{\rm Quantitative comparisons of different distribution strategies.}
    \begin{tabular}{|c|c|c|c|c|c|c|}
    \cline{1-7}
      &  \multicolumn{3}{c|}{Noise-free data training} & \multicolumn{3}{c|}{Noisy data Tuning ($n = 10^3$)}\\
    \cline{1-7}
    Photons&  \multicolumn{1}{c|}{$n = 10^5$}&  \multicolumn{1}{c|}{$n = 10^3$} & \multicolumn{1}{c|}{$n = 10^2$} & \multicolumn{1}{c|}{$n = 10^5$}&  \multicolumn{1}{c|}{$n = 10^3$} & \multicolumn{1}{c|}{$n = 10^2$}\\
    \cline{1-7}
      Criteria&PSNR / SSIM&PSNR / SSIM&PSNR / SSIM&PSNR / SSIM&PSNR / SSIM&PSNR / SSIM\\
    \cline{1-7}
    Strategy 1  & 28.63 / 0.8742 & 27.32 / 0.8462 & 25.98 / 0.8082 &
                  27.96 / 0.8614 & 27.71 / 0.8566 & 26.44 / 0.8233 \\
    \cline{1-7}
    Strategy 2  & 27.19 / 0.8448 & 27.05 / 0.8408 & 26.23 / 0.8142 &
                  27.19 / 0.8446 & 27.12 / 0.8434 & 26.43 / 0.8222 \\
    Strategy 3  & 27.63 / 0.8533 & 27.28 / 0.8453 & 25.60 / 0.7913 & 
                  27.59 / 0.8512 & 27.56 / 0.8506 & 25.97 / 0.8040 \\
    Strategy 4  & 28.32 / 0.8691 & 27.74 / 0.8564 & 25.70 / 0.7900 &
                  28.16 / 0.8663 & 28.02 / 0.8633 & 25.88 / 0.7991 \\
    \cline{1-7}
    Strategy 5  & 26.59 / 0.8300 & 26.23 / 0.8208 & 24.97 / 0.7799 &
                  26.66 / 0.8279 & 26.60 / 0.8263 & 25.36 / 0.7901 \\
    Strategy 6  & 24.66 / 0.7665 & 24.48 / 0.7618 & 23.87 / 0.7463 &
                  24.64 / 0.7667 & 24.69 / 0.7674 & 24.12 / 0.7526 \\
    \cline{1-7}
    Strategy 7  & 27.56 / 0.8556 & 27.22 / 0.8483 & 25.80 / 0.8090 &
                  27.55 / 0.8546 & 27.42 / 0.8518 & 26.08 / 0.8107 \\
    Strategy 8  & 27.79 / 0.8577 & 27.49 / 0.8508 & 26.05 / 0.8053 & 
                  27.88 / 0.8563 & 27.75 / 0.8541 & 26.53 / 0.8211 \\
    Strategy 9  & 27.84 / 0.8586 & 27.48 / 0.8507 & 25.99 / 0.8029 & 
                  27.94 / 0.8573 & 27.79 / 0.8546 & 26.41 / 0.8144 \\
    Strategy 10 & 27.07 / 0.8384 & 26.97 / 0.8367 & 26.36 / 0.8202 & 
                  27.14 / 0.8387 & 27.07 / 0.8375 & 26.46 / 0.8221 \\
    \cline{1-7}
    \end{tabular}%
    \label{tab:simulation_compare_sparse}%
\end{table}%
    
Results above and in the supplement document show that sparse sampling strategies can produce visually satisfactory reconstruction for most images. Further improving the reconstruction fidelity requires increasing the number of measured pixels and the Fourier space coverage. This process is easily achieved within a camera array system. Because the system throughput has been increased with multiple cameras, the reconstruction fidelity can be significantly improved by spatially shifting the system and increasing the number of snapshots. We simulate this process following the aperture distribution strategy 4. 

In this given aperture distribution, each snapshot captures 9 images, covering 44\% of the Fourier space. To cover more Fourier space, we can shift the system and have multiple snapshots, so in total $9k$ images are captured with $k$ snapshots. We simulated $k = 1...6$, increasing the covered Fourier space from 44\% to 100\%. In terms of reconstruction network, we increased the growth rate to 24 and the number of building blocks in each dense block to 6. With the increased number of measurements and coverage of the Fourier space, the reconstruction may be further improved with traditional alternating projection algorithms by initializing the algorithms with the network prediction. As a benchmark, we simulated the traditional FP with standard 61\% overlap and 100 measurements. We also simulated applying only alternating projection algorithm on the 54 images captured from 6 snapshots. For fair comparison, we maintained the total number of photons from the source image, so the number of photons per measurement decreased with the increased overlap in Fourier space. We show the simulation results in Table \ref{tab:simulation_compare_fp}. Image samples and the detailed aperture distributions are provided in Section 3 in Supplement 1.

The results show that: 1) in low noise conditions, increasing the measurements will improve the reconstruction fidelity. With alternating projection algorithm, the system achieves performance competitive to traditional FP; 2) with the increased noise level, the deep learning method outperforms the alternating projection with much fewer measurements.

Another aspect of this is that the proposed system does not require accurate calibration and a forward model, so, instead of shifting the camera system, the shift in Fourier space can be achieved by adjusting the illumination angle. One can consider adding extra illumination sources or installing the laser on a translator.

\begin{table}[htbp]
\footnotesize
  \centering
  \caption{\rm Quantitative comparisons of different number of snapshots. The percentages of the measured Fourier space are labeled in the table.}
    \begin{tabular}{|c|c|c|c|c|c|c|}
    \cline{1-7}
      &  \multicolumn{3}{c|}{NN prediction} & \multicolumn{3}{c|}{NN prediction + Alternating projection}\\
    \cline{1-7}
    Photons&  \multicolumn{1}{c|}{$n = 10^5$}&  \multicolumn{1}{c|}{$n = 10^3$} & \multicolumn{1}{c|}{$n = 10^2$} & \multicolumn{1}{c|}{$n = 10^5$}&  \multicolumn{1}{c|}{$n = 10^3$} & \multicolumn{1}{c|}{$n = 10^2$}\\
    \cline{1-7}
      Criteria&PSNR / SSIM&PSNR / SSIM&PSNR / SSIM&PSNR / SSIM&PSNR / SSIM&PSNR / SSIM\\
    \cline{1-7}
    1 snapshot (44.3\%) & 28.43 / 0.8688 & 27.84 / 0.8568 & 25.75 / 0.7945&
                   -&-& -\\
    2 snapshots (69.7\%)& 31.23 / 0.9250 & 29.68 / 0.9020 & 26.29 / 0.8295 &
                  31.47 / 0.9253 & 28.88 / 0.8821 & 24.54 / 0.7357\\
    3 snapshots (85.7\%)& 32.92 / 0.9475 & 30.74 / 0.9196 & 26.90 / 0.8415 &
                  34.46 / 0.9578 & 29.79 / 0.9003 & 25.04 / 0.7594\\
    4 snapshots (95.4\%)& 33.81 / 0.9578 & 31.17 / 0.9260 & 26.95 / 0.8425&
                  37.38 / 0.9773 & 30.49 / 0.9122 & 25.10 / 0.7580\\
    5 snapshots (98.7\%) & 33.96 / 0.9600 & 30.97 / 0.9236 & 26.66 / 0.8320&
                 40.05 / 0.9877 & 30.99 / 0.9208 & 25.96 / 0.7974\\
    6 snapshots (100\%) & 34.14 / 0.9620 & 31.07 / 0.9267 & 26.68 / 0.8376&
                  42.38 / 0.9928& 31.35 / 0.9253 & 25.94 / 0.7940 \\
    \hline
    \end{tabular}
    \vspace{4mm}

    \begin{tabular}{|c|c|c|c|}
    \cline{1-4}
    \multicolumn{4}{|c|}{Alternating Projection}\\
    \cline{1-4}
    Photons&  \multicolumn{1}{c|}{$n = 10^5$}&  \multicolumn{1}{c|}{$n = 10^3$} & \multicolumn{1}{c|}{$n = 10^2$}\\
    \cline{1-4}
    Criteria&PSNR / SSIM&PSNR / SSIM&PSNR / SSIM\\
    \cline{1-4}
    FP & 46.01 / 0.9967 & 31.10 / 0.9203& 25.30 / 0.7686\\
    \cline{1-4}
    6 snapshots& 32.15 / 0.9371 & 29.40 / 0.8959& 25.67 / 0.7867\\
    \cline{1-4}
    \end{tabular}%
  \label{tab:simulation_compare_fp}%
\end{table}%

\section{Conclusion}
\label{sec:conclusion}
We have shown that it is possible to combine coherent image data over multiple camera apertures to super-resolve a remote scene with a single snapshot of data. Of course, our system is contrived in the sense that we have full control over the object field through an SLM, which allows us to train the system without fully calibrating the structure of the forward model. In future work, we hope to build on the results presented here to create synthetic aperture images of natural objects. We imagine that such an imaging system can be calibrated with a combination of structured illumination and test objects, but we leave demonstration of such calibration to future work. We have also compared diverse array structures and found that unstructured arrays perform best with snapshot reconstruction. Again referring to future work, we anticipate that multiframe estimation over moving platforms will further improve these results. 

\bibliographystyle{plain}
\bibliography{references}

\begin{thebibliography}{10}

\bibitem{Agustsson_2017_CVPR_Workshops}
Eirikur Agustsson and Radu Timofte.
\newblock Ntire 2017 challenge on single image super-resolution: Dataset and
  study.
\newblock In {\em The IEEE Conference on Computer Vision and Pattern
  Recognition (CVPR) Workshops}, July 2017.

\bibitem{beck2005synthetic}
Steven~M Beck, Joseph~R Buck, Walter~F Buell, Richard~P Dickinson, David~A
  Kozlowski, Nicholas~J Marechal, and Timothy~J Wright.
\newblock Synthetic-aperture imaging laser radar: laboratory demonstration and
  signal processing.
\newblock {\em Applied optics}, 44(35):7621--7629, 2005.

\bibitem{bian2014content}
Liheng Bian, Jinli Suo, Guohai Situ, Guoan Zheng, Feng Chen, and Qionghai Dai.
\newblock Content adaptive illumination for fourier ptychography.
\newblock {\em Optics letters}, 39(23):6648--6651, 2014.

\bibitem{boominathan2018phase}
Lokesh Boominathan, Mayug Maniparambil, Honey Gupta, Rahul Baburajan, and
  Kaushik Mitra.
\newblock Phase retrieval for fourier ptychography under varying amount of
  measurements.
\newblock {\em arXiv preprint arXiv:1805.03593}, 2018.

\bibitem{brady2009optical}
David~J Brady.
\newblock {\em Optical imaging and spectroscopy}.
\newblock John Wiley \& Sons, 2009.

\bibitem{brady2011gigapixel}
David~J Brady and Sehoon Lim.
\newblock Gigapixel holography.
\newblock In {\em 2011 ICO International Conference on Information Photonics},
  pages 1--2. IEEE, 2011.

\bibitem{chan2019parallel}
Antony~CS Chan, Jinho Kim, An~Pan, Han Xu, Dana Nojima, Christopher Hale,
  Songli Wang, and Changhuei Yang.
\newblock Parallel fourier ptychographic microscopy for high-throughput
  screening with 96 cameras (96 eyes).
\newblock {\em Scientific reports}, 9(1):1--12, 2019.

\bibitem{chung2016wide}
Jaebum Chung, Jinho Kim, Xiaoze Ou, Roarke Horstmeyer, and Changhuei Yang.
\newblock Wide field-of-view fluorescence image deconvolution with
  aberration-estimation from fourier ptychography.
\newblock {\em Biomedical optics express}, 7(2):352--368, 2016.

\bibitem{clic}
CLIC.
\newblock Workshop and challenge on learned image compression, 2018.

\bibitem{dong2014fpscope}
Siyuan Dong, Kaikai Guo, Pariksheet Nanda, Radhika Shiradkar, and Guoan Zheng.
\newblock Fpscope: a field-portable high-resolution microscope using a
  cellphone lens.
\newblock {\em Biomedical optics express}, 5(10):3305--3310, 2014.

\bibitem{dong2014aperture}
Siyuan Dong, Roarke Horstmeyer, Radhika Shiradkar, Kaikai Guo, Xiaoze Ou,
  Zichao Bian, Huolin Xin, and Guoan Zheng.
\newblock Aperture-scanning fourier ptychography for 3d refocusing and
  super-resolution macroscopic imaging.
\newblock {\em Optics express}, 22(11):13586--13599, 2014.

\bibitem{fienup2011gigapixel}
James~R Fienup and Abbie~E Tippie.
\newblock Gigapixel synthetic-aperture digital holography.
\newblock In {\em Tribute to Joseph W. Goodman}, volume 8122, page 812203.
  International Society for Optics and Photonics, 2011.

\bibitem{gharbi2016deep}
Micha{\"e}l Gharbi, Gaurav Chaurasia, Sylvain Paris, and Fr{\'e}do Durand.
\newblock Deep joint demosaicking and denoising.
\newblock {\em ACM Transactions on Graphics (ToG)}, 35(6):1--12, 2016.

\bibitem{Guo_2020_CVPR}
Yong Guo, Jian Chen, Jingdong Wang, Qi~Chen, Jiezhang Cao, Zeshuai Deng, Yanwu
  Xu, and Mingkui Tan.
\newblock Closed-loop matters: Dual regression networks for single image
  super-resolution.
\newblock In {\em Proceedings of the IEEE/CVF Conference on Computer Vision and
  Pattern Recognition (CVPR)}, June 2020.

\bibitem{he2018single}
Xiaoliang He, Cheng Liu, and Jianqiang Zhu.
\newblock Single-shot fourier ptychography based on diffractive beam splitting.
\newblock {\em Optics letters}, 43(2):214--217, 2018.

\bibitem{holloway2016toward}
Jason Holloway, M~Salman Asif, Manoj~Kumar Sharma, Nathan Matsuda, Roarke
  Horstmeyer, Oliver Cossairt, and Ashok Veeraraghavan.
\newblock Toward long-distance subdiffraction imaging using coherent camera
  arrays.
\newblock {\em IEEE Transactions on Computational Imaging}, 2(3):251--265,
  2016.

\bibitem{holloway2017savi}
Jason Holloway, Yicheng Wu, Manoj~K Sharma, Oliver Cossairt, and Ashok
  Veeraraghavan.
\newblock Savi: Synthetic apertures for long-range, subdiffraction-limited
  visible imaging using fourier ptychography.
\newblock {\em Science advances}, 3(4):e1602564, 2017.

\bibitem{horstmeyer2016diffraction}
Roarke Horstmeyer, Jaebum Chung, Xiaoze Ou, Guoan Zheng, and Changhuei Yang.
\newblock Diffraction tomography with fourier ptychography.
\newblock {\em Optica}, 3(8):827--835, 2016.

\bibitem{huang2017densely}
Gao Huang, Zhuang Liu, Laurens Van Der~Maaten, and Kilian~Q Weinberger.
\newblock Densely connected convolutional networks.
\newblock In {\em Proceedings of the IEEE conference on computer vision and
  pattern recognition}, pages 4700--4708, 2017.

\bibitem{jiang2018solving}
Shaowei Jiang, Kaikai Guo, Jun Liao, and Guoan Zheng.
\newblock Solving fourier ptychographic imaging problems via neural network
  modeling and tensorflow.
\newblock {\em Biomedical optics express}, 9(7):3306--3319, 2018.

\bibitem{kappeler2017ptychnet}
Armin Kappeler, Sushobhan Ghosh, Jason Holloway, Oliver Cossairt, and Aggelos
  Katsaggelos.
\newblock Ptychnet: Cnn based fourier ptychography.
\newblock In {\em 2017 IEEE International Conference on Image Processing
  (ICIP)}, pages 1712--1716. IEEE, 2017.

\bibitem{kim2016incubator}
Jinho Kim, Beverley~M Henley, Charlene~H Kim, Henry~A Lester, and Changhuei
  Yang.
\newblock Incubator embedded cell culture imaging system (emsight) based on
  fourier ptychographic microscopy.
\newblock {\em Biomedical optics express}, 7(8):3097--3110, 2016.

\bibitem{kingma2014adam}
Diederik~P Kingma and Jimmy Ba.
\newblock Adam: A method for stochastic optimization.
\newblock {\em arXiv preprint arXiv:1412.6980}, 2014.

\bibitem{konda2020fourier}
Pavan~Chandra Konda, Lars Loetgering, Kevin~C Zhou, Shiqi Xu, Andrew~R Harvey,
  and Roarke Horstmeyer.
\newblock Fourier ptychography: current applications and future promises.
\newblock {\em Optics express}, 28(7):9603--9630, 2020.

\bibitem{konda2021multi}
Pavan~Chandra Konda, Jonathan~M Taylor, and Andrew~R Harvey.
\newblock Multi-aperture fourier ptychographic microscopy, theory and
  validation.
\newblock {\em Optics and Lasers in Engineering}, 138:106410, 2021.

\bibitem{krause2011synthetic}
Brian~W Krause, Joe Buck, Chris Ryan, David Hwang, Piotr Kondratko, Andrew
  Malm, Andy Gleason, and Shaun Ashby.
\newblock Synthetic aperture ladar flight demonstration.
\newblock In {\em CLEO: Science and Innovations}, page PDPB7. Optical Society
  of America, 2011.

\bibitem{lee2018single}
Byounghyo Lee, Jong-young Hong, Dongheon Yoo, Jaebum Cho, Youngmo Jeong, Seokil
  Moon, and Byoungho Lee.
\newblock Single-shot phase retrieval via fourier ptychographic microscopy.
\newblock {\em Optica}, 5(8):976--983, 2018.

\bibitem{nguyen2018deep}
Thanh Nguyen, Yujia Xue, Yunzhe Li, Lei Tian, and George Nehmetallah.
\newblock Deep learning approach for fourier ptychography microscopy.
\newblock {\em Optics express}, 26(20):26470--26484, 2018.

\bibitem{ou2016aperture}
Xiaoze Ou, Jaebum Chung, Roarke Horstmeyer, and Changhuei Yang.
\newblock Aperture scanning fourier ptychographic microscopy.
\newblock {\em Biomedical Optics Express}, 7(8):3140--3150, 2016.

\bibitem{ou2013quantitative}
Xiaoze Ou, Roarke Horstmeyer, Changhuei Yang, and Guoan Zheng.
\newblock Quantitative phase imaging via fourier ptychographic microscopy.
\newblock {\em Optics letters}, 38(22):4845--4848, 2013.

\bibitem{ou2015high}
Xiaoze Ou, Roarke Horstmeyer, Guoan Zheng, and Changhuei Yang.
\newblock High numerical aperture fourier ptychography: principle,
  implementation and characterization.
\newblock {\em Optics express}, 23(3):3472--3491, 2015.

\bibitem{persson2012reducing}
Martin Persson, David Engstr{\"o}m, and Mattias Goks{\"o}r.
\newblock Reducing the effect of pixel crosstalk in phase only spatial light
  modulators.
\newblock {\em Optics express}, 20(20):22334--22343, 2012.

\bibitem{ronneberger2015u}
Olaf Ronneberger, Philipp Fischer, and Thomas Brox.
\newblock U-net: Convolutional networks for biomedical image segmentation.
\newblock In {\em International Conference on Medical image computing and
  computer-assisted intervention}, pages 234--241. Springer, 2015.

\bibitem{ryle1960synthesis}
Martin Ryle and Anthony Hewish.
\newblock The synthesis of large radio telescopes.
\newblock {\em Monthly Notices of the Royal Astronomical Society},
  120(3):220--230, 1960.

\bibitem{schulz2021photon}
Timothy~J Schulz, David~J Brady, and Chengyu Wang.
\newblock Photon-limited bounds for phase retrieval.
\newblock {\em Optics Express}, 29(11):16736--16748, 2021.

\bibitem{shamshad2019deep}
Fahad Shamshad, Farwa Abbas, and Ali Ahmed.
\newblock Deep ptych: Subsampled fourier ptychography using generative priors.
\newblock In {\em ICASSP 2019-2019 IEEE International Conference on Acoustics,
  Speech and Signal Processing (ICASSP)}, pages 7720--7724. IEEE, 2019.

\bibitem{sun2018single}
Jiasong Sun, Qian Chen, Jialin Zhang, Yao Fan, and Chao Zuo.
\newblock Single-shot quantitative phase microscopy based on color-multiplexed
  fourier ptychography.
\newblock {\em Optics letters}, 43(14):3365--3368, 2018.

\bibitem{wang2018inverse}
Ning Wang, Ran Wang, Di~Mo, Guangzuo Li, Keshu Zhang, and Yirong Wu.
\newblock Inverse synthetic aperture ladar demonstration: system structure,
  imaging processing, and experiment result.
\newblock {\em Applied optics}, 57(2):230--236, 2018.

\bibitem{code}
{Wang, C., Hu, M., and Brady, D. J.}
\newblock Snapshot ptychography code, 2021.
\newblock
  \url{https://github.com/djbradyAtOpticalSciencesArizona/arrayCameraFourierPtychography}.

\bibitem{xue2019reliable}
Yujia Xue, Shiyi Cheng, Yunzhe Li, and Lei Tian.
\newblock Reliable deep-learning-based phase imaging with uncertainty
  quantification.
\newblock {\em Optica}, 6(5):618--629, 2019.

\bibitem{yeh2015experimental}
Li-Hao Yeh, Jonathan Dong, Jingshan Zhong, Lei Tian, Michael Chen, Gongguo
  Tang, Mahdi Soltanolkotabi, and Laura Waller.
\newblock Experimental robustness of fourier ptychography phase retrieval
  algorithms.
\newblock {\em Optics express}, 23(26):33214--33240, 2015.

\bibitem{yuan2021snapshot}
Xin Yuan, David~J Brady, and Aggelos~K Katsaggelos.
\newblock Snapshot compressive imaging: Theory, algorithms, and applications.
\newblock {\em IEEE Signal Processing Magazine}, 38(2):65--88, 2021.

\bibitem{zaperty2018numerical}
Weronika Zaperty and Tomasz Kozacki.
\newblock Numerical model of diffraction effects of pixelated phase-only
  spatial light modulators.
\newblock In {\em Speckle 2018: VII International Conference on Speckle
  Metrology}, volume 10834, page 108342A. International Society for Optics and
  Photonics, 2018.

\bibitem{zhang2019fourier}
Jizhou Zhang, Tingfa Xu, Ziyi Shen, Yifan Qiao, and Yizhou Zhang.
\newblock Fourier ptychographic microscopy reconstruction with multiscale deep
  residual network.
\newblock {\em Optics express}, 27(6):8612--8625, 2019.

\bibitem{zhang2017beyond}
Kai Zhang, Wangmeng Zuo, Yunjin Chen, Deyu Meng, and Lei Zhang.
\newblock Beyond a gaussian denoiser: Residual learning of deep cnn for image
  denoising.
\newblock {\em IEEE transactions on image processing}, 26(7):3142--3155, 2017.

\bibitem{zhang2014fundamentals}
Zichen Zhang, Zheng You, and Daping Chu.
\newblock Fundamentals of phase-only liquid crystal on silicon (lcos) devices.
\newblock {\em Light: Science \& Applications}, 3(10):e213--e213, 2014.

\bibitem{zheng2016fourier}
Guoan Zheng.
\newblock {\em Fourier ptychographic imaging: a MATLAB tutorial}.
\newblock Morgan \& Claypool Publishers, 2016.

\bibitem{zheng2013wide}
Guoan Zheng, Roarke Horstmeyer, and Changhuei Yang.
\newblock Wide-field, high-resolution fourier ptychographic microscopy.
\newblock {\em Nature photonics}, 7(9):739--745, 2013.

\bibitem{zheng2021concept}
Guoan Zheng, Cheng Shen, Shaowei Jiang, Pengming Song, and Changhuei Yang.
\newblock Concept, implementations and applications of fourier ptychography.
\newblock {\em Nature Reviews Physics}, 3(3):207--223, 2021.

\end{thebibliography}

\newpage
\section*{Supplement Document}
\setcounter{section}{0}
\setcounter{figure}{0}

This supplemental document includes reconstruction samples from the physical system along with their measurements, reconstruction samples from simulations with various aperture sizes and distributions, visualization of the multi-snapshot process and the corresponding reconstruction samples.

\section{Visual results of the Physical Setup}
Fig.~\ref{fig:slm_full1} and Fig.~\ref{fig:slm_full2} show samples from our physical setup of array-camera snapshot Fourier ptychography (FP). For each sample, on the left we show the network output, the thresholded output, the ground truth image and the image directly down-sampled from ground truth, and on the right we show the 16 measured images from the camera array. The camera array does not directly measure the target, and both the structural and the textural information are estimated by the neural network from the measured interference between pixels.

\newcommand{\figm}{0.243}
\newcommand{\drawmeasurement}[1]{
	\includegraphics[width=\figm\linewidth]{#1/measure_1.png}
	\includegraphics[width=\figm\linewidth]{#1/measure_2.png}
	\includegraphics[width=\figm\linewidth]{#1/measure_3.png}
	\includegraphics[width=\figm\linewidth]{#1/measure_4.png} \\
	\includegraphics[width=\figm\linewidth]{#1/measure_5.png}
	\includegraphics[width=\figm\linewidth]{#1/measure_6.png} 
	\includegraphics[width=\figm\linewidth]{#1/measure_7.png}
	\includegraphics[width=\figm\linewidth]{#1/measure_8.png} \\
	\includegraphics[width=\figm\linewidth]{#1/measure_9.png}
	\includegraphics[width=\figm\linewidth]{#1/measure_10.png}
	\includegraphics[width=\figm\linewidth]{#1/measure_11.png}
	\includegraphics[width=\figm\linewidth]{#1/measure_12.png}\\
	\includegraphics[width=\figm\linewidth]{#1/measure_13.png}
	\includegraphics[width=\figm\linewidth]{#1/measure_14.png}
	\includegraphics[width=\figm\linewidth]{#1/measure_15.png}
	\includegraphics[width=\figm\linewidth]{#1/measure_16.png}\\
}

\newcommand{\drawslm}[1]{
\scriptsize
\begin{minipage}[b]{0.402\linewidth}
\centering
	\stackunder[1pt]{\includegraphics[width=0.49\linewidth]{#1/output.png}}{Network output}
	\stackunder[1pt]{\includegraphics[width=0.49\linewidth]{#1/output_thresh.png}}{Reconstruction} 
	\stackunder[1pt]{\includegraphics[width=0.49\linewidth]{#1/label.png}}{Ground truth} 
	\stackunder[1pt]{\includegraphics[width=0.49\linewidth]{#1/downsample.png}}{Down-sampled} \\
\end{minipage}
\hspace{3mm}
\begin{minipage}[b]{0.43\linewidth}
\centering
\drawmeasurement{#1}
\end{minipage}}

\begin{figure}[htbp]
\centering
    \drawslm{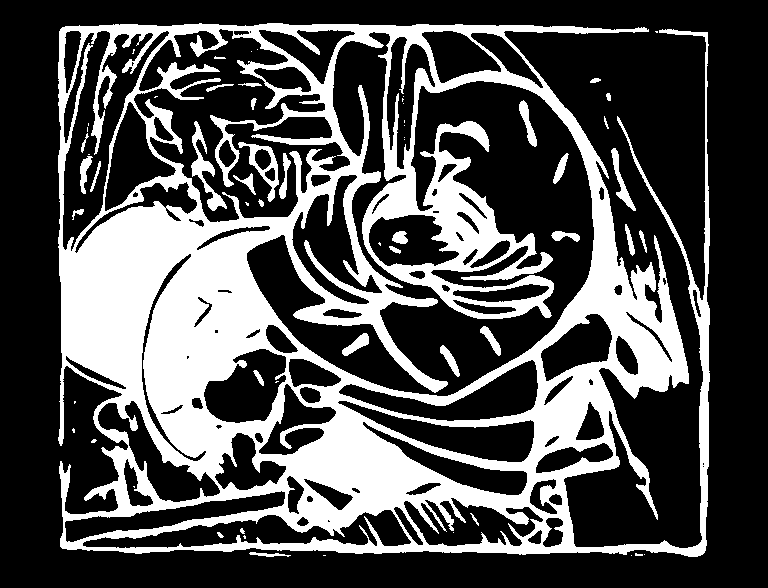}\\
    \vspace{0.5mm}
    \drawslm{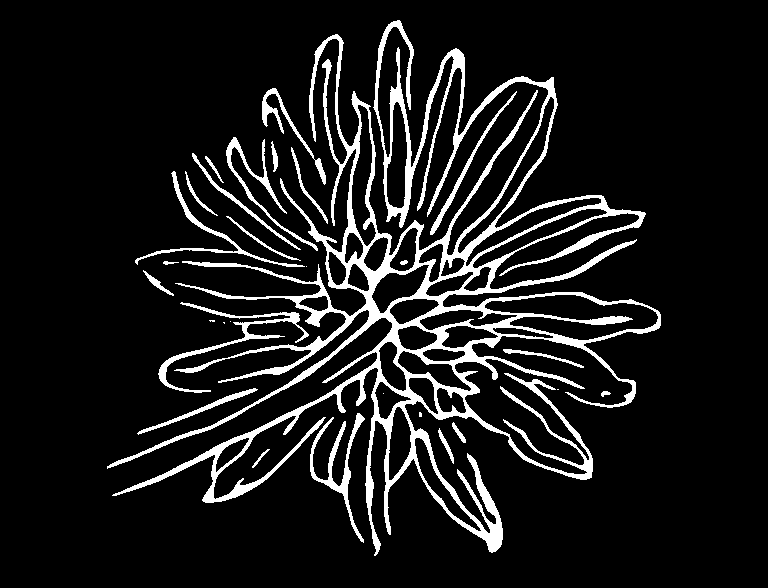}
\caption{\label{fig:slm_full1}Visual results of the array-camera snapshot FP. The image intensity of the measurements has been adjusted for visualization.}
\end{figure}

\begin{figure}[htbp]
\centering
    \drawslm{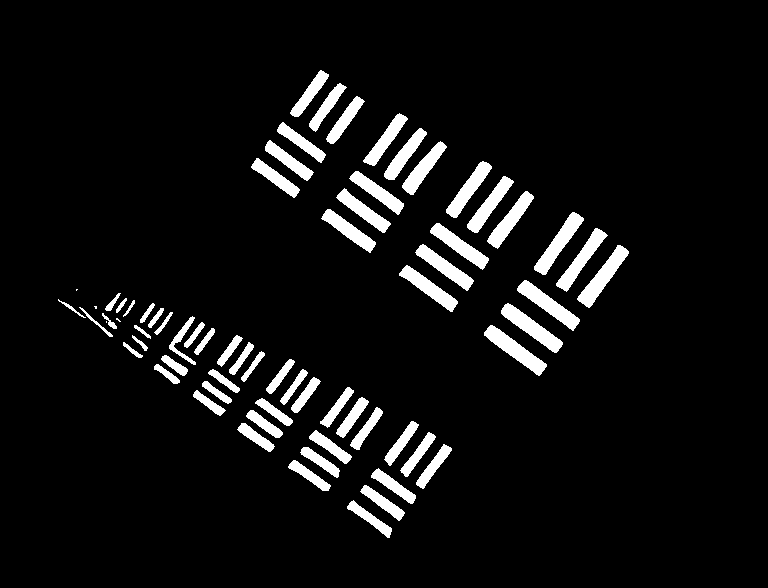}\\
    \vspace{0.5mm}
    \drawslm{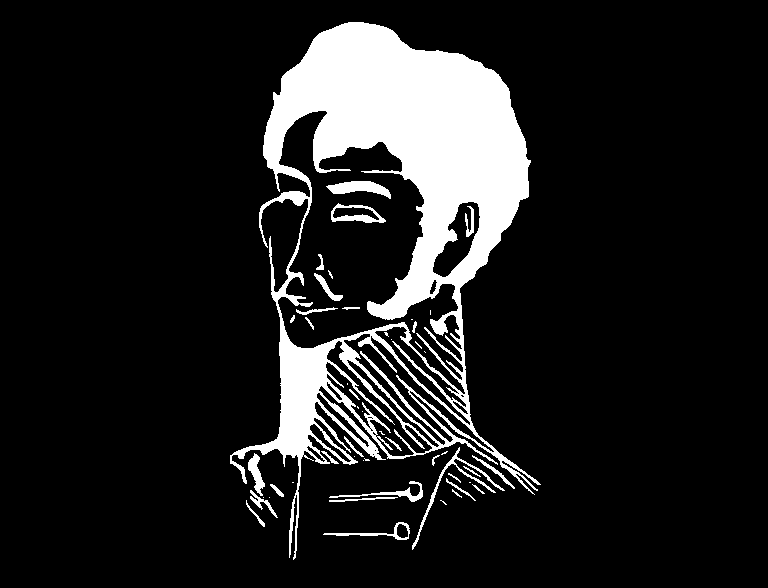}\\
    \vspace{0.5mm}
    \drawslm{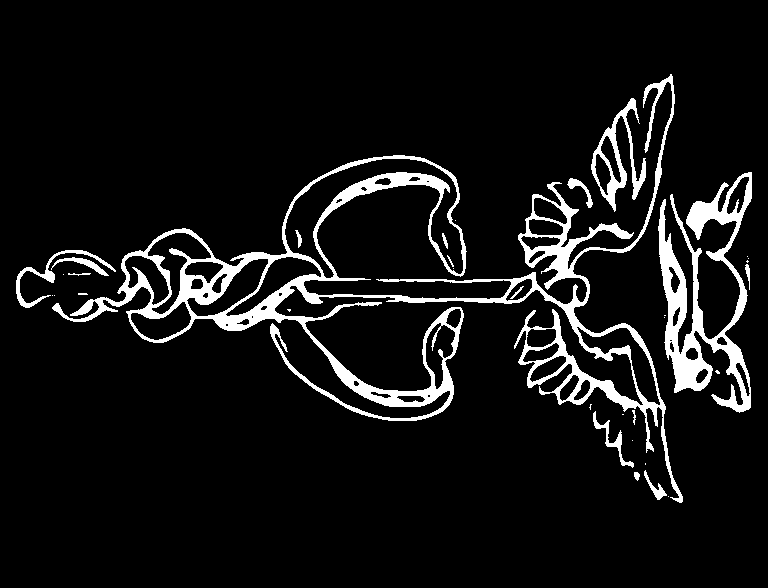}\\
    \vspace{0.5mm}
    \drawslm{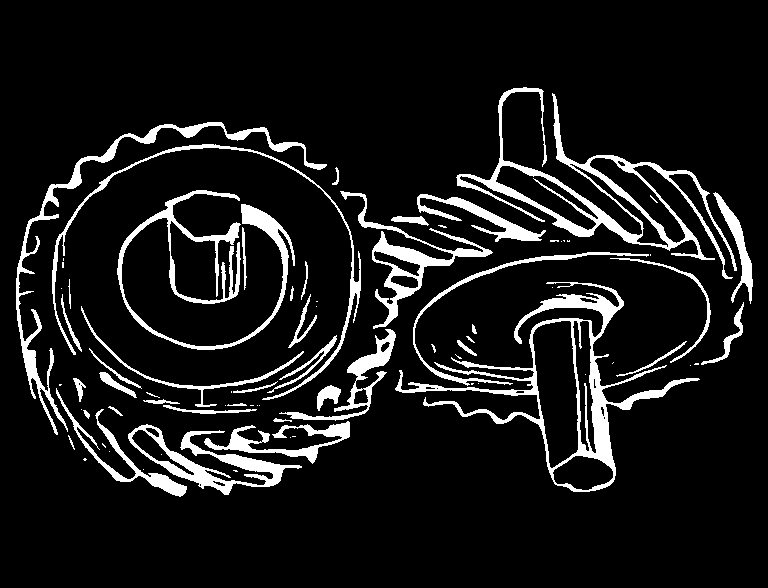}
\caption{\label{fig:slm_full2}Visual results of the array-camera snapshot FP. The image intensity of the measurements has been adjusted for visualization.}
\end{figure}

\section{Visual analysis of the impact of subaperture size and distribution}
We compare the visual results of the reconstructed images with various subaperture sizes and distributions in simulation. Fig.~\ref{fig:sup_full_img} illustrates 10 different strategies. Each source image contains $512\times512$ pixels, and each measured image contains $128\times128$, $96\times96$ or $64\times64$ pixels. See the paper for more details. Three reconstructed images are shown in Fig.~\ref{fig:sup_full_img}, and five zoomed-in details are shown in Fig.~\ref{fig:sup_img_detail}. Compared with densely distributed and uniformly distributed apertures, sparsely and randomly distributed apertures best preserve high-frequency texture information. Further comparisons also show that combining multi-scale apertures achieves competitive results, which allows using smaller and cheaper lenses in building the system.

\section{Visual analysis of different number of snapshots}
\begin{figure}[htbp]
\centering
\subfigure[\label{fig:distribution_sup}1 snapshot, 44.3\%]{\includegraphics[width=0.25\textwidth]{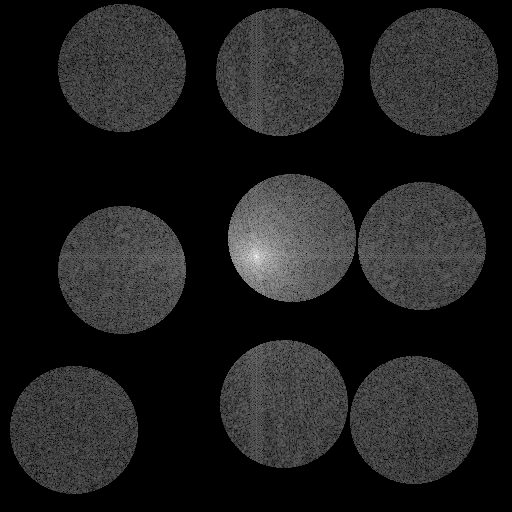}}
\hspace{1mm}
\subfigure[2 snapshots, 69.7\%]{\includegraphics[width=0.25\textwidth]{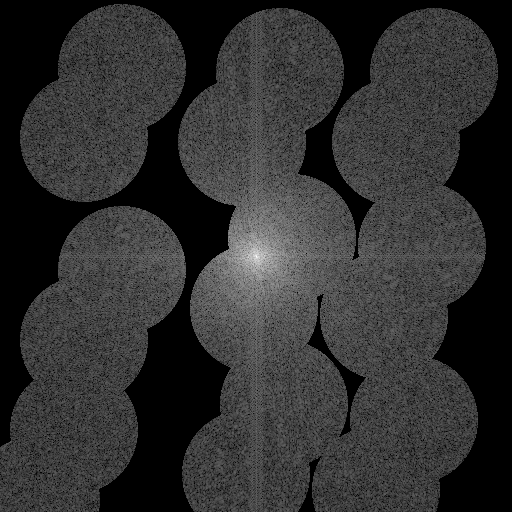}}
\hspace{1mm}
\subfigure[3 snapshots, 85.7\%]{\includegraphics[width=0.25\textwidth]{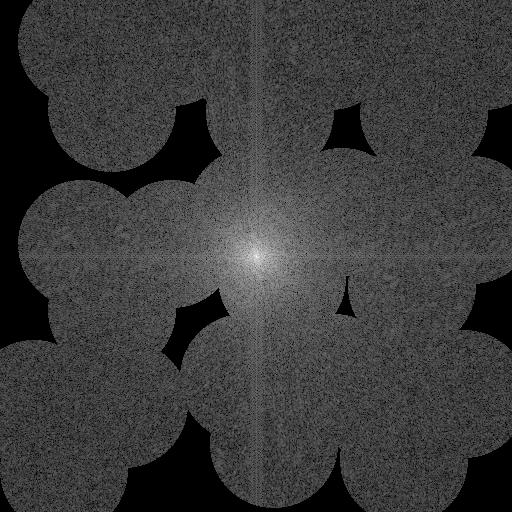}}
\hspace{1mm}
\subfigure[4 snapshots, 95.4\%]{\includegraphics[width=0.25\textwidth]{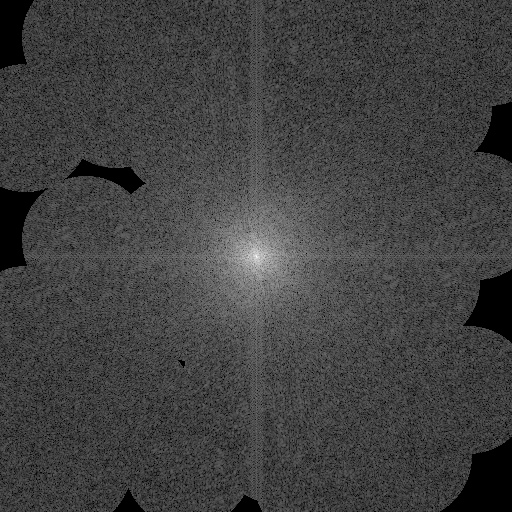}}
\hspace{1mm}
\subfigure[5 snapshots, 98.7\%]{\includegraphics[width=0.25\textwidth]{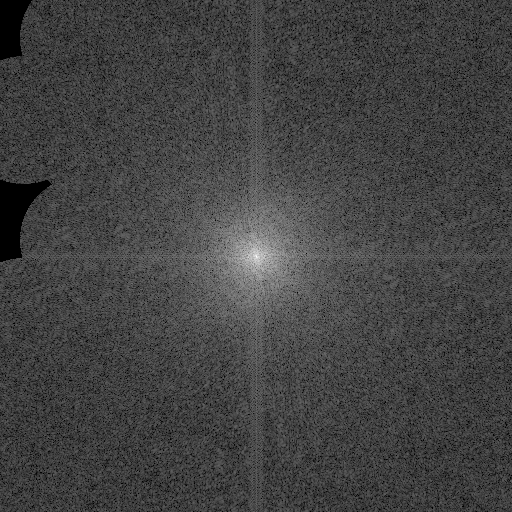}}
\hspace{1mm}
\subfigure[6 snapshots, 100\%]{\includegraphics[width=0.25\textwidth]{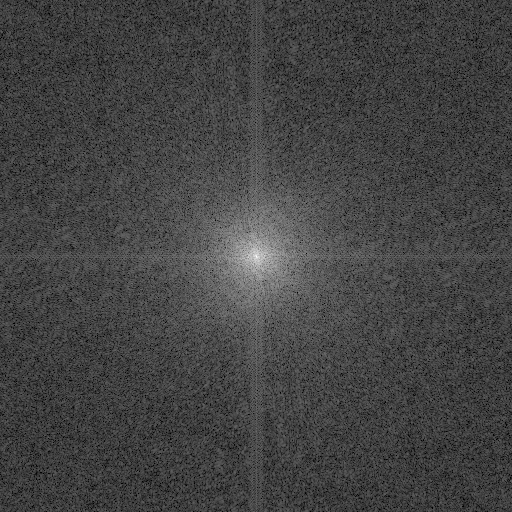}}
\caption{\label{fig:sup_multiple_exps}Fourier space coverage with different number of snapshots. The percentage of the covered Fourier space increases from 44.3\% to 100\%. }
\end{figure}

The reconstruction fidelity can be improved by spatially shifting the system and increasing the number of snapshots. We simulate this process using the aperture distribution in Fig.~\ref{fig:distribution_sup}. After each snapshot, we shifted the apertures before next snapshot to cover more Fourier space. We can cover the entire Fourier space with 6 snapshots. The increasing coverage is shown in Fig.~\ref{fig:sup_multiple_exps}. We show reconstruction samples in Fig.~\ref{fig:sup_compare_multi_exps}. Because the system throughput has been increased with the camera array, we see 3 dB improvement in PSNR by adding only one more snapshot. When the entire Fourier space is covered, the network achieves 6 dB improvement in PSNR.

\newcommand{\figw}{0.150}
\newcommand{\drawfiguresup}[1]{
\scriptsize
    \stackunder[1pt]{\includegraphics[width=\figw\linewidth]{#1/gt.png}}{Ground truth} 
    \hspace{1.5mm}
	\stackunder[1pt]{\includegraphics[width=\figw\linewidth]{#1/1.png}}{Strategy 1}  
	\stackunder[1pt]{\includegraphics[width=\figw\linewidth]{#1/2.png}}{Strategy 2}  
	\stackunder[1pt]{\includegraphics[width=\figw\linewidth]{#1/3.png}}{Strategy 3}
	\stackunder[1pt]{\includegraphics[width=\figw\linewidth]{#1/4.png}}{Strategy 4} 
	\stackunder[1pt]{\includegraphics[width=\figw\linewidth]{#1/5.png}}{Strategy 5}
	\vspace{0.6mm}
	\stackunder[1pt]{\includegraphics[width=\figw\linewidth]{#1/bicubic.png}}{Interpolation}
	\hspace{1.5mm}
	\stackunder[1pt]{\includegraphics[width=\figw\linewidth]{#1/6.png}}{Strategy 6} 
	\stackunder[1pt]{\includegraphics[width=\figw\linewidth]{#1/7.png}}{Strategy 7}
	\stackunder[1pt]{\includegraphics[width=\figw\linewidth]{#1/8.png}}{Strategy 8} 
	\stackunder[1pt]{\includegraphics[width=\figw\linewidth]{#1/9.png}}{Strategy 9}
	\stackunder[1pt]{\includegraphics[width=\figw\linewidth]{#1/10.png}}{Strategy 10} 
}

\newcommand{\drawdistribution}[1]{
\scriptsize
    \stackunder[1pt]{\includegraphics[width=\figw\linewidth]{#1/gt.png}}{Ground truth} 
    \hspace{1.5mm}
	\stackunder[1pt]{\includegraphics[width=\figw\linewidth]{#1/1.png}}{Strategy 1}  
	\stackunder[1pt]{\includegraphics[width=\figw\linewidth]{#1/2.png}}{Strategy 2}  
	\stackunder[1pt]{\includegraphics[width=\figw\linewidth]{#1/3.png}}{Strategy 3}
	\stackunder[1pt]{\includegraphics[width=\figw\linewidth]{#1/4.png}}{Strategy 4} 
	\stackunder[1pt]{\includegraphics[width=\figw\linewidth]{#1/5.png}}{Strategy 5}
	\vspace{0.4mm}
	\stackunder[1pt]{\includegraphics[width=\figw\linewidth]{#1/fourier.png}}{Fourier Space}
	\hspace{1.5mm}
	\stackunder[1pt]{\includegraphics[width=\figw\linewidth]{#1/6.png}}{Strategy 6} 
	\stackunder[1pt]{\includegraphics[width=\figw\linewidth]{#1/7.png}}{Strategy 7}
	\stackunder[1pt]{\includegraphics[width=\figw\linewidth]{#1/8.png}}{Strategy 8} 
	\stackunder[1pt]{\includegraphics[width=\figw\linewidth]{#1/9.png}}{Strategy 9}
	\stackunder[1pt]{\includegraphics[width=\figw\linewidth]{#1/10.png}}{Strategy 10} 
}

\newcommand{\figwdsup}{0.147}
\newcommand{\drawfiguredetail}[1]{
\centering
\scriptsize
    \stackunder[1pt]{\includegraphics[width=\figwdsup\linewidth]{#1/gt.png}}{Ground truth} 
    \hspace{0.4mm}
	\stackunder[1pt]{\includegraphics[width=\figwdsup\linewidth]{#1/1.png}}{Strategy 1}  
	\stackunder[1pt]{\includegraphics[width=\figwdsup\linewidth]{#1/2.png}}{Strategy 2}  
	\stackunder[1pt]{\includegraphics[width=\figwdsup\linewidth]{#1/3.png}}{Strategy 3}
	\stackunder[1pt]{\includegraphics[width=\figwdsup\linewidth]{#1/4.png}}{Strategy 4} 
	\stackunder[1pt]{\includegraphics[width=\figwdsup\linewidth]{#1/5.png}}{Strategy 5}\\

	\stackunder[1pt]{\includegraphics[width=\figwdsup\linewidth]{#1/bicubic.png}}{Interpolation}
	\hspace{0.4mm}
	\stackunder[1pt]{\includegraphics[width=\figwdsup\linewidth]{#1/6.png}}{Strategy 6} 
	\stackunder[1pt]{\includegraphics[width=\figwdsup\linewidth]{#1/7.png}}{Strategy 7}
	\stackunder[1pt]{\includegraphics[width=\figwdsup\linewidth]{#1/8.png}}{Strategy 8} 
	\stackunder[1pt]{\includegraphics[width=\figwdsup\linewidth]{#1/9.png}}{Strategy 9}
	\stackunder[1pt]{\includegraphics[width=\figwdsup\linewidth]{#1/10.png}}{Strategy 10} \\
}

\newcommand{\drawdetail}[1]{
\scriptsize
\hspace{5mm}
\begin{minipage}[b]{0.14\linewidth}
\centering
	\raisebox{+0.5\height}{\stackunder[1pt]{\includegraphics[width=\linewidth]{#1/image.png}}{Ground truth} }
\end{minipage}
% \hspace{0.1mm}
\begin{minipage}[b]{0.8\linewidth}
\centering
\drawfiguredetail{#1}
\end{minipage}}

\begin{figure}[htbp]
\centering
    \drawdistribution{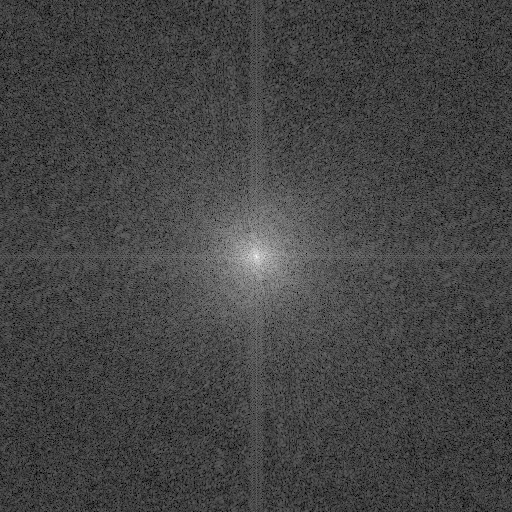}\\
    \begin{tikzpicture}
    \draw[dashed,thick] (-7.6,0) -- (7.6,0);
    \end{tikzpicture}
	\vspace{1mm}\\
    \drawfiguresup{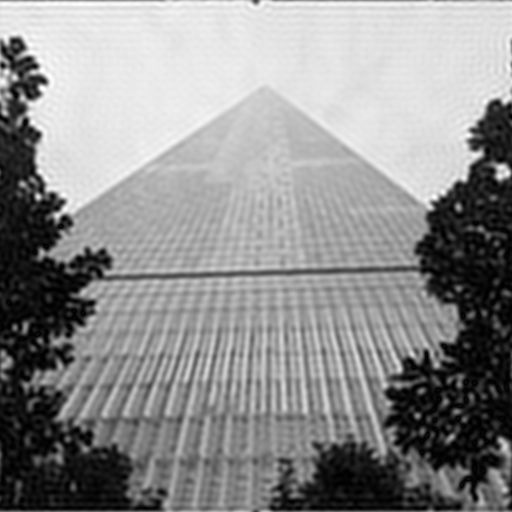}\\
	\vspace{1mm}
    \drawfiguresup{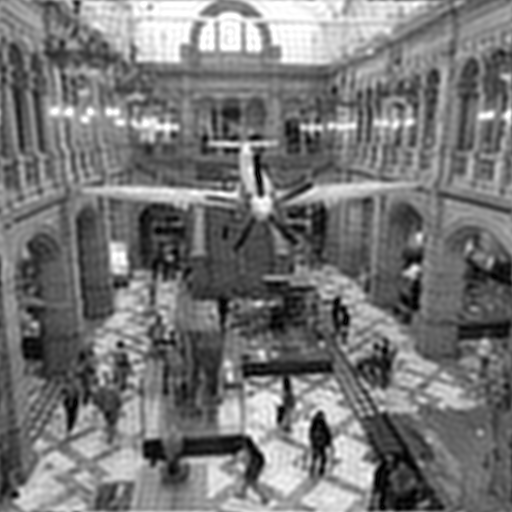}\\
	\vspace{1mm}
    \drawfiguresup{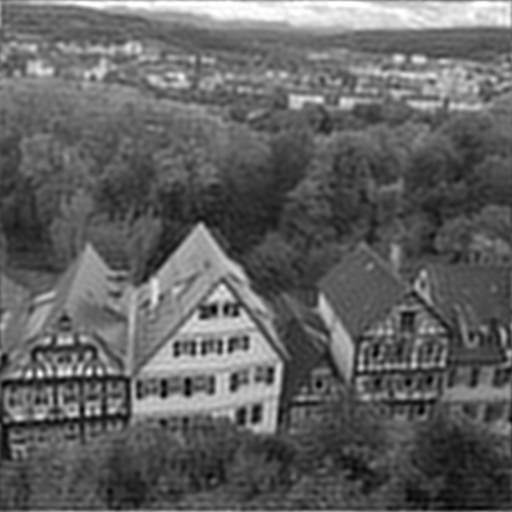} 
\caption{\label{fig:sup_full_img}Visual comparison between different aperture sizes and distributions. Zoom in to see image details.}
\end{figure}

\begin{figure}[ht]
\centering
    \drawdetail{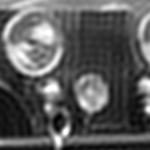}\\
    \vspace{1mm}
    \drawdetail{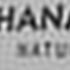}\\
    \vspace{1mm}
    \drawdetail{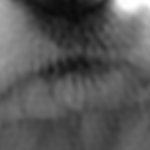}\\
    \vspace{1mm}
    \drawdetail{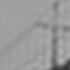}\\
    \vspace{1mm}
    \drawdetail{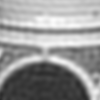}
    \caption{\label{fig:sup_img_detail}Visual comparison between different aperture sizes and distributions. }
\end{figure}

\newcolumntype{C}{>{\centering\arraybackslash}m{8.8em}}
\begin{figure}%\sffamily
\centering
\begin{tabular}{c@{\hspace{5pt}}C@{\hspace{3pt}}C@{\hspace{3pt}}C@{\hspace{3pt}}C}

% \toprule
% Exps\\ %& a & b & c & d \\ 
% \toprule
    \scriptsize 1 snapshot& \includegraphics[width=8.8em]{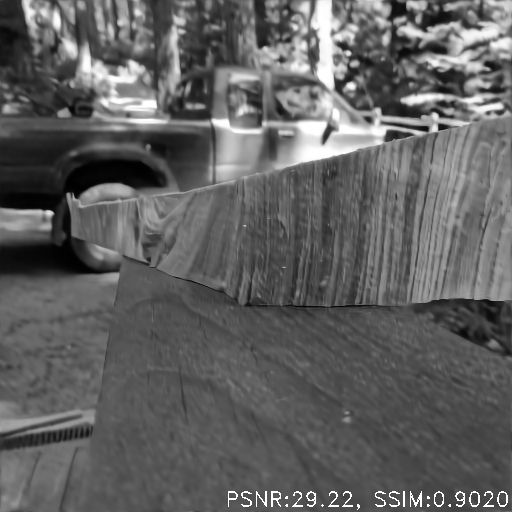} &  \includegraphics[width=8.8em]{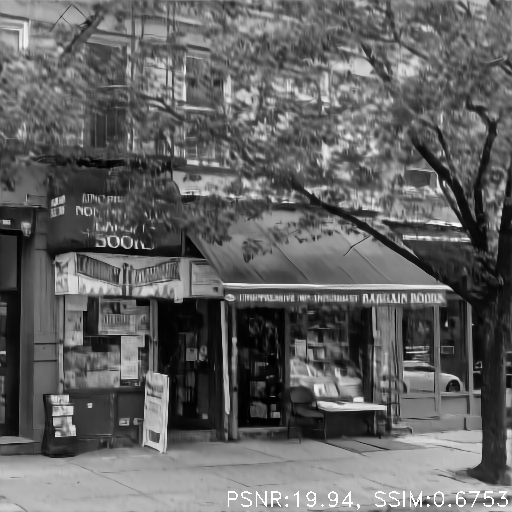} & \includegraphics[width=8.8em]{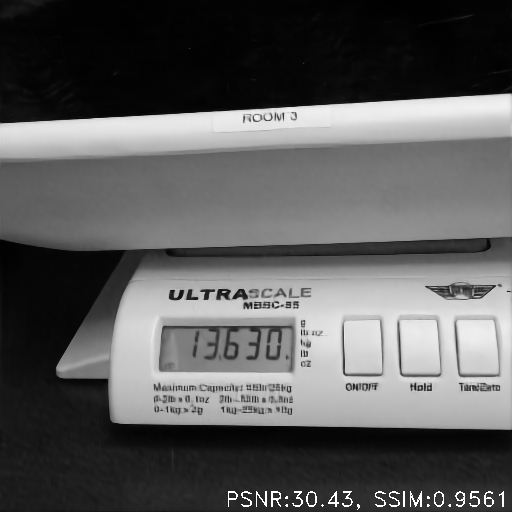} & \includegraphics[width=8.8em]{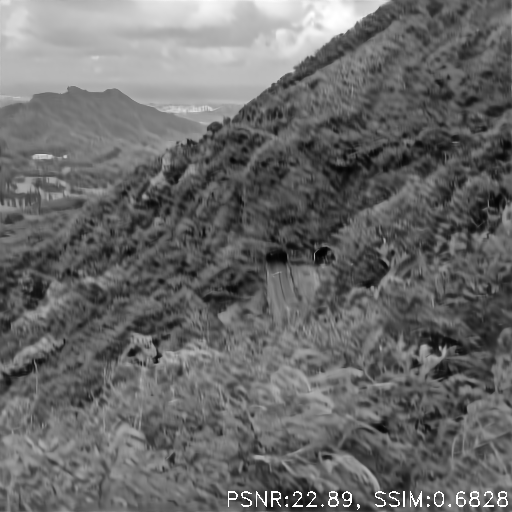}  \\ 
    \scriptsize 2 snapshots& \includegraphics[width=8.8em]{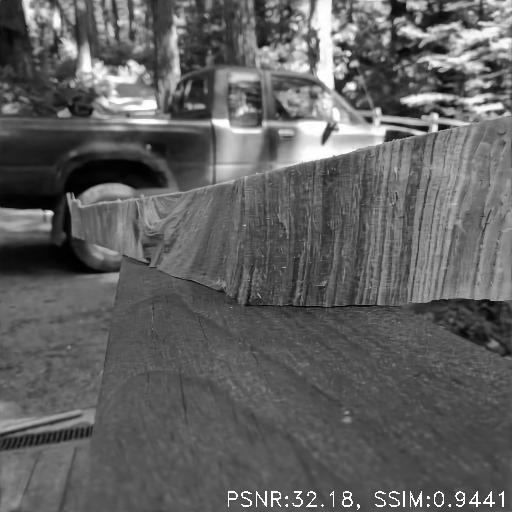} &  \includegraphics[width=8.8em]{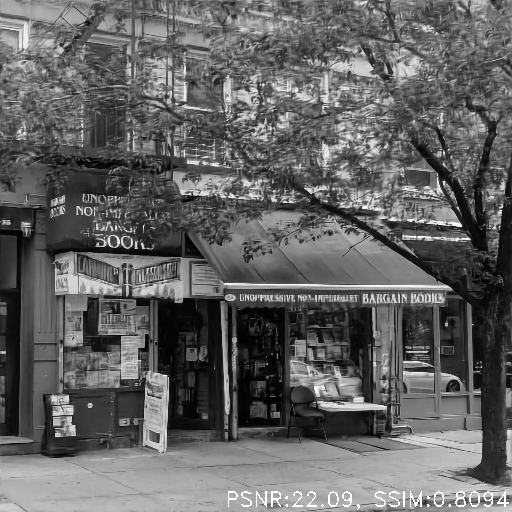} & \includegraphics[width=8.8em]{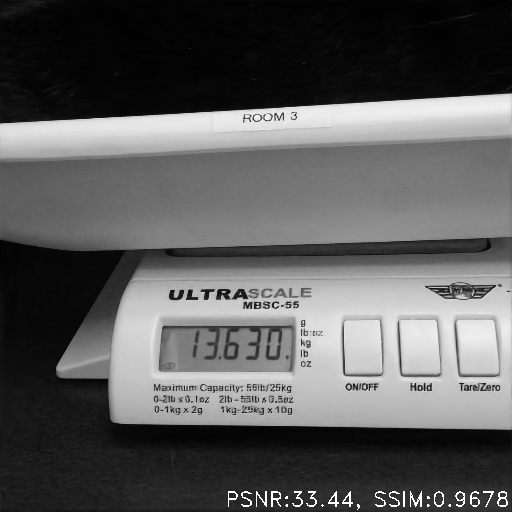} & \includegraphics[width=8.8em]{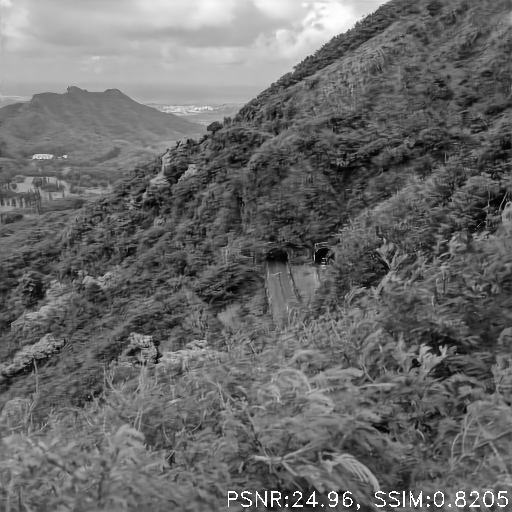}  \\  
    \scriptsize 3 snapshots& \includegraphics[width=8.8em]{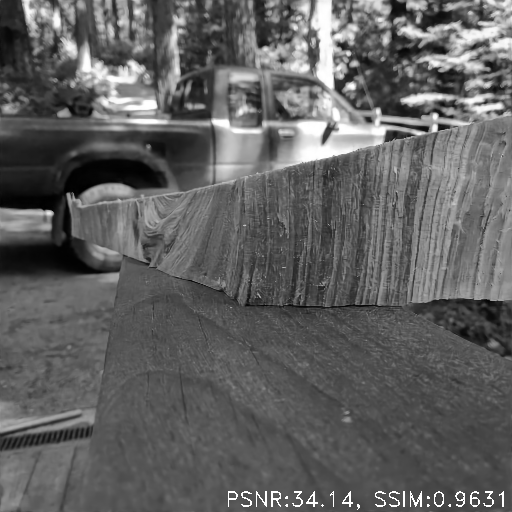} &  \includegraphics[width=8.8em]{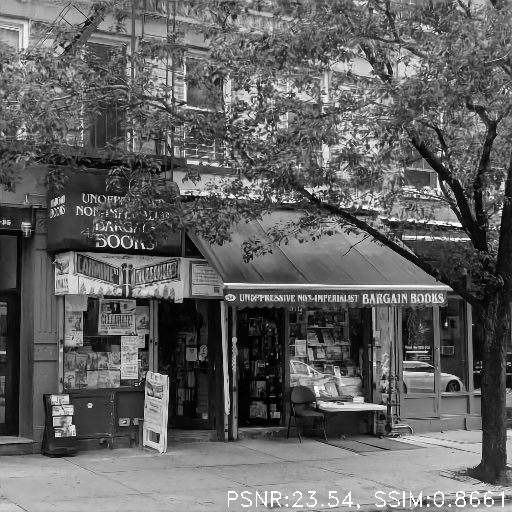} & \includegraphics[width=8.8em]{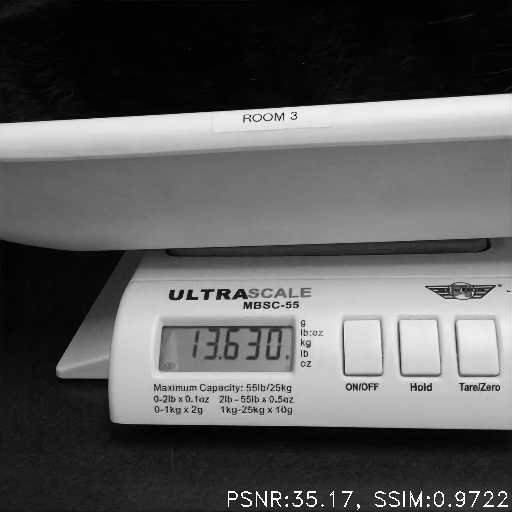} & \includegraphics[width=8.8em]{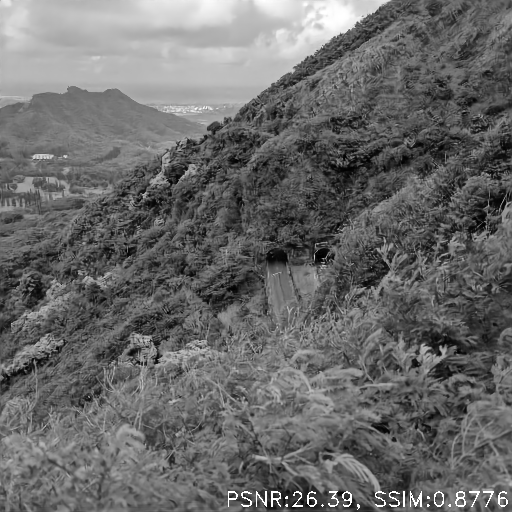}  \\ 
    \scriptsize 4 snapshots& \includegraphics[width=8.8em]{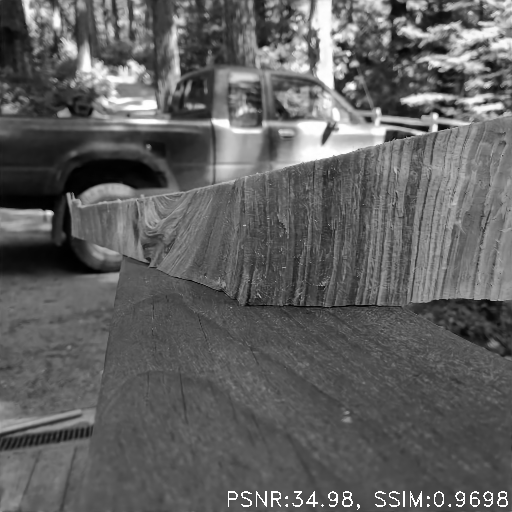} &  \includegraphics[width=8.8em]{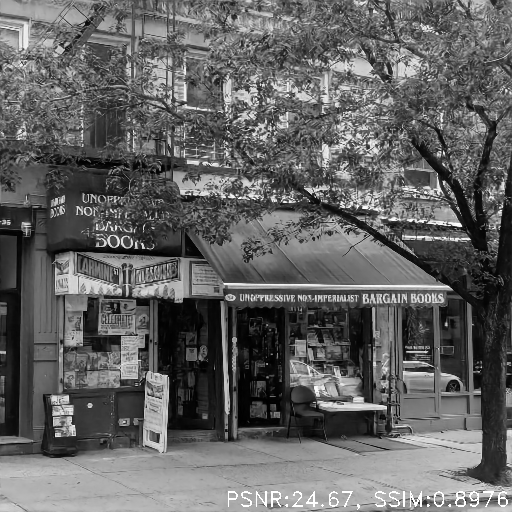} & \includegraphics[width=8.8em]{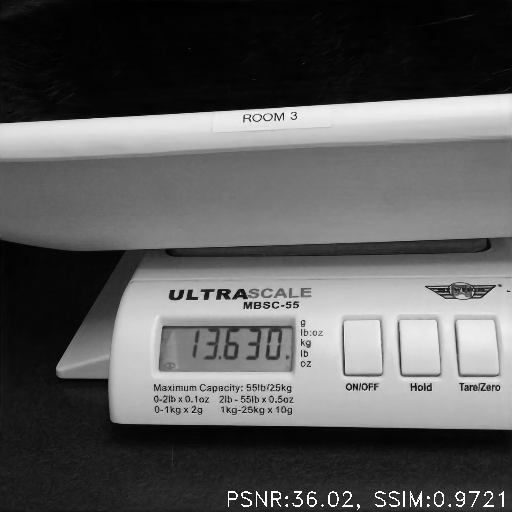} & \includegraphics[width=8.8em]{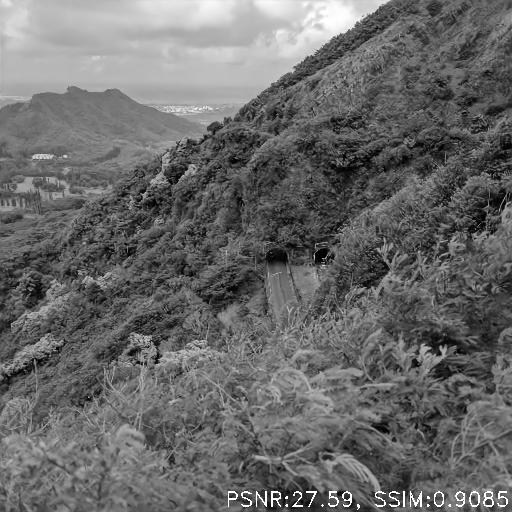} \\ 
    \scriptsize 5 snapshots& \includegraphics[width=8.8em]{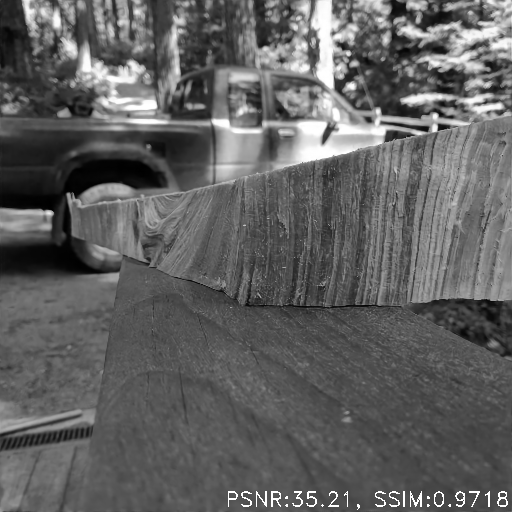} &  \includegraphics[width=8.8em]{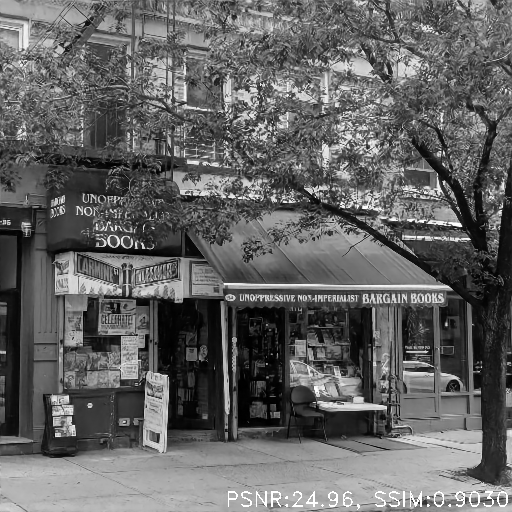} & \includegraphics[width=8.8em]{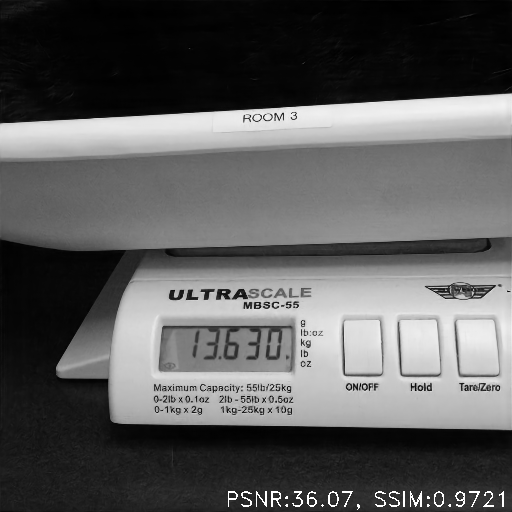} & \includegraphics[width=8.8em]{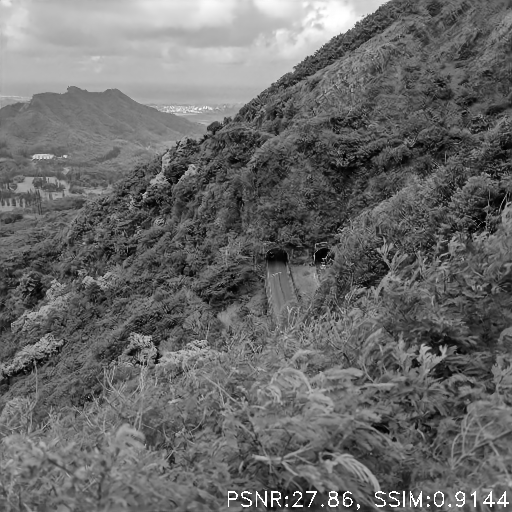} \\ 
    \scriptsize 6 snapshots& \includegraphics[width=8.8em]{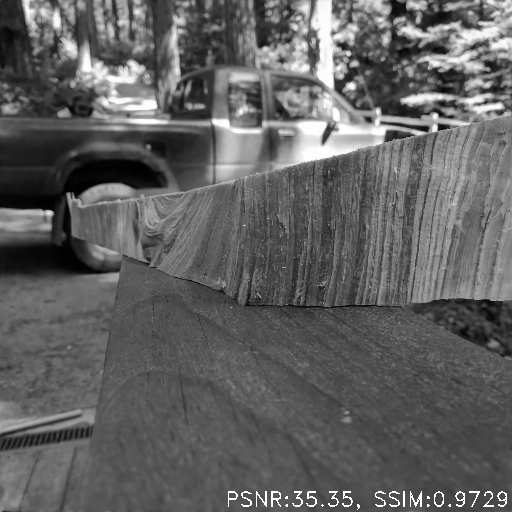} &  \includegraphics[width=8.8em]{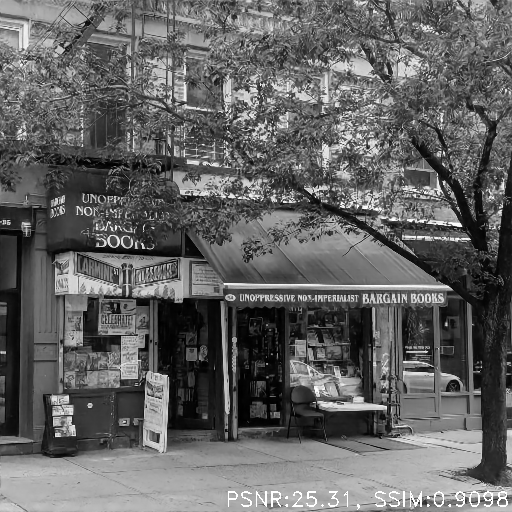} & \includegraphics[width=8.8em]{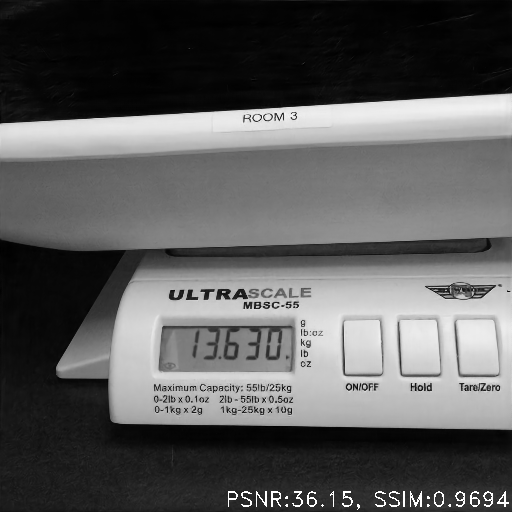} & \includegraphics[width=8.8em]{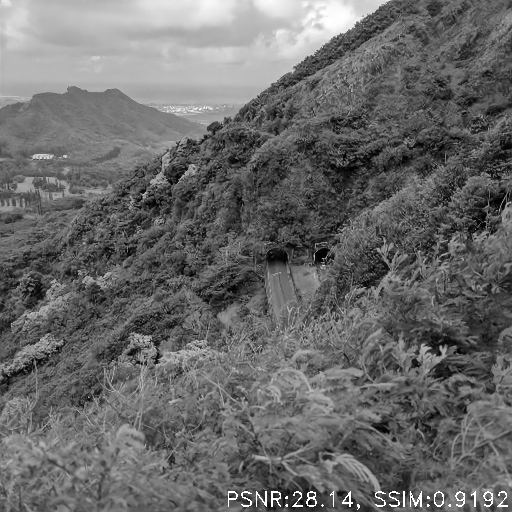} \\ 
    \scriptsize Ground truth & \includegraphics[width=8.8em]{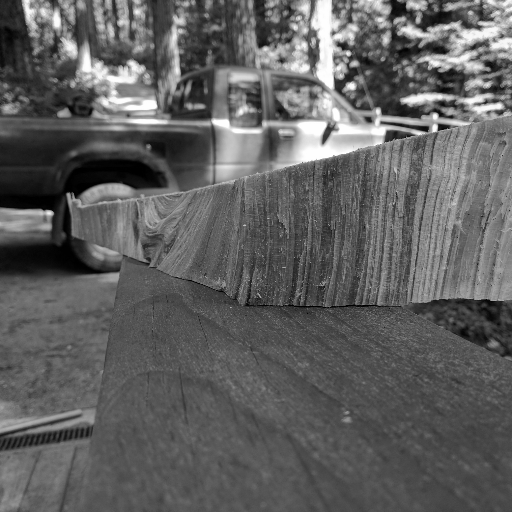}&  \includegraphics[width=8.8em]{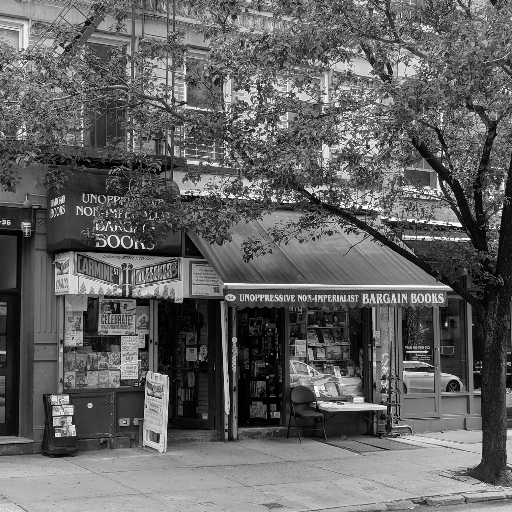}& \includegraphics[width=8.8em]{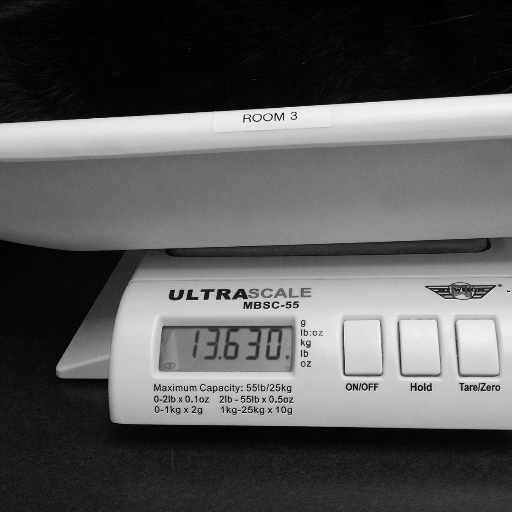}&
    \includegraphics[width=8.8em]{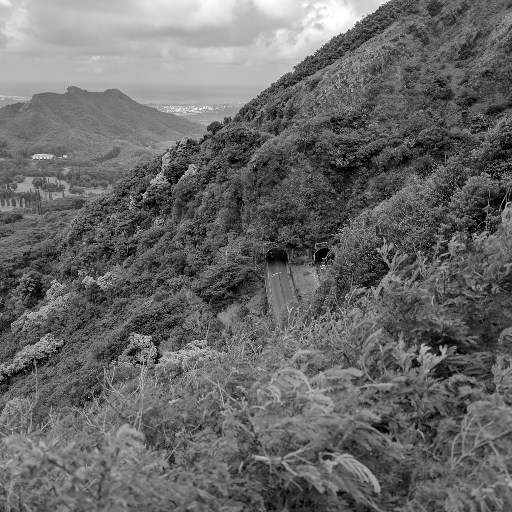}\\ 
% \bottomrule 
\end{tabular}
\caption{\label{fig:sup_compare_multi_exps}Visual results with different number of snapshots.}
\end{figure} 
\end{document}